\documentclass[11pt,psfig]{article}
\usepackage{graphicx}
\usepackage{fancyheadings}
\usepackage{setspace}               
\include{definitions}               
\usepackage{amsfonts}
\usepackage{amsmath}
\usepackage{amssymb}
\usepackage{amsthm}
\usepackage{caption}
\usepackage{subcaption}
\usepackage{titlesec}
\usepackage{xcolor}
\usepackage[left=2cm,right=2cm,top=2cm,bottom=2cm]{geometry}

\DeclareGraphicsExtensions{.png,.pdf}

\newtheorem{theorem}{Theorem}[section]

\newtheorem{proposition}[theorem]{Proposition}
\newtheorem{corollary}[theorem]{Corollary}

\theoremstyle{definition}

\newtheorem{definition}[theorem]{Definition}
\theoremstyle{remark}
\newtheorem{remark}[theorem]{Remark}
\newtheorem{assumption}[theorem]{Assumption}

\numberwithin{equation}{section}

\def\sP{\mathbb{P}}

\def\sR{\mathbb{R}}
\def\sR{\mathbb{R}}

\def\sE{\mathbb{E}}

\newcommand{\Ac}{\mathcal{A}}

\newcommand{\g}{\gamma}

\newcommand{\be}{\begin{equation}}
\newcommand{\ee}{\end{equation}}
\newcommand{\bea}{\begin{eqnarray}}
\newcommand{\eea}{\end{eqnarray}}
\newcommand{\bean}{\begin{eqnarray*}}
\newcommand{\eean}{\end{eqnarray*}}
\def \proof{{\noindent \bf Proof}\quad}
\def \ep{\hbox{ }\hfill$\Box$}

\makeatletter 
\@addtoreset{equation}{section}

\makeatother 
\begin{document}

\title{Portfolio Time Consistency and Utility Weighted Discount Rates }

\author{Oumar Mbodji\thanks{%
Independent Researcher}\\
\texttt{oumarsoule@gmail.com}
\and Traian A. Pirvu\thanks{%
Department of Mathematics and Statistics, McMaster University, 1280 Main
Street West, Hamilton, ON, L8S 4K1, Canada} \\
\texttt{tpirvu@math.mcmaster.ca} }
\maketitle

\noindent \textbf{Abstract.} 
 Merton portfolio management problem is studied in this paper within a stochastic volatility, non constant time discount rate, and power utility framework.
This problem is time inconsistent and the way out of this predicament is to consider the subgame perfect strategies. The
later are characterized through  an extended Hamilton Jacobi Bellman (HJB) equation. A fixed point iteration is employed to solve the extended HJB equation. This is done in a two stage approach: in a first
step the utility weighted discount rate is introduced and characterized as the fixed point of a certain operator; in the second
step the value function is determined through a linear parabolic partial differential equation. Numerical experiments
explore the effect of the time discount rate on the subgame perfect and precommitment strategies.
\addcontentsline{toc}{section}{Abstract}


\vspace*{7mm}


\vspace*{5mm}


 \section{Introduction}
 
This paper studies a time inconsistent stochastic control problem, the Merton portfolio management within stochastic volatility and non constant time discount 
 rate setting. The risk preferences are of power type. By now, there is substantial evidence that people discount the future outcomes at a
non-constant rate. More precisely, there is experimental evidence (see
 \cite{Fr_Lo_Te1} for a review) that people are more sensitive
to a given time delay if it occurs earlier. Individual behaviour is perhaps best described by 
\emph{hyperbolic discounting}, where the discount function is $h(t)=(1+at)^{-%
\frac{b}{a}},$ with $a, b>0$. The corresponding discount rate is $\rho(
t) =\frac{b}{ 1+at} $, which starts from $\rho(0) = b $ and decreases to zero. Because of its empirical support, hyperbolic
discounting has received a lot of attention in the areas of: microeconomics,
macroeconomics and behavioural finance. 

 The Merton portfolio management problem with non constant time discount rate is known to be time inconsistent, meaning that optimal strategies computed
 in the past fail to remain optimal. Let us provide a review on this topic. There is a vast literature by now on time inconsistent stochastic control portfolio problems. We only present some of the works. The first paper to consider the Merton problem with non-constant discount rates is  \cite{Ek_Pi1}; they show the problem is time inconsistent, and introduce the subgame perfect strategies 
as Nash equilibrium time consistent strategies. In a follow up paper \cite{Ek_Mb_Pi1} characterizes the subgame perfect strategies in terms of a certain
"value function", which is shown to satisfy an extended HJB equation. The  value function of a more general stochastic control problem is shown to solve an extended HJB equation in \cite{Bj_Kh_Mu1}. The time inconsistent dynamic mean-variance portfolio problem is solved in \cite{Ba2}, and the time consistent subgame perfect strategy is found by means of dynamic programming. The work of \cite{Yo1} considers the well posedness of the extended HJB when the diffusion term of the state process does not depend on the control, and it shows the existence of a subgame perfect strategy. The wealth dependent risk aversion dynamic mean-variance portfolio problem is addressed in \cite{Bj_Mu_Zh1}, and a time consistent strategy is found. The time inconsistent portfolio problem when market coefficients and discount factor switch according to a finite state Markov chain is solved in \cite{Pi_Zh1}. The paper \cite{Yo2} studies a time inconsistent stochastic optimal control problem with a recursive cost functional, and an equilibrium strategy is characterized by means of multiperson hierarchical differential games. A new notion of time consistent strategies, the regular equilibrium strategies, are introduced by \cite{He}, and they are compared to
the subgame perfect (Nash equilibrium) strategies. In a highly tractable continuous-time Markov chain model \cite{H7} employes a fixed point argument to determine subgame perfect strategies.

There are recent studies on time inconsistent optimal stopping problems. The subgame perfect strategies are found by fixed point iterations; see \cite{H1} for 
stopping under non-exponential discounting, and \cite{H3}, \cite{H5}, \cite{H6} for more general settings. There are extensions of  \cite{H1} to incorporate  probability distortions and model ambiguity, see
\cite{H2}, \cite{H4}.

 The goal of this paper is to study the difference between 
 subgame perfect  investment strategy and the precommitment  one; the later is the optimal investment strategy computed at current time when the agent pre commits to it. There is no difference between these two investment strategies when the stock has constant drift and volatility. Thus, the stock's drift and volatility are stochastic in our framework.

The methodology developed in \cite{Ek_Mb_Pi1} is employed here
to find the subgame perfect strategies through a value function, which is characterized by an extended HJB equation. The novelty of our approach is the \emph{the utility weighted discount rate} a strategy dependent quantity. In the special case of exponential discounting this equals the constant (psychological) discount rate. 
The utility weighted discount rate of the subgame perfect strategies admits a fixed point characterization. This comes to support
the economic intuition that subgame perfect strategies are in fact fixed points of an interpersonal game. We show, using a decoupling technique, that the utility weighted discount rate induced by the subgame perfect strategy is
exogenously determined. This finding makes it possible to reduce the generalized (extended) HJB to a linear parabolic PDE. 
Our methodology yields a recipe to find subgame perfect strategies when the interest rate, the market price of risk
 and the stock volatility are well behaved e.g. adapted to the Brownian filtration, bounded and sufficiently regular.
A numerical scheme based on fixed point iterations makes it possible to visualize the subgame perfect strategies as well as the precommitment strategies.


\textbf{Contributions}: Let us summarize the findings and contributions of our paper. The subgame perfect strategies are characterized through a value function
which solves an extended HJB (a system of PDEs, SDE and an integral equation with a non local term). The utility weighted discount rate, a strategy dependent quantity, 
it is shown to be a fixed point of an operator, and this reduces the extended HJB to a parabolic linear equation. A novel approach of finding subgame perfect strategies 
through fixed point iterations is introduced and implemented. The study of utility weighted discount rate reveals some properties of the subgame perfect strategies. The precommitment  investment strategy turns out to be independent of the discount function. On the other hand the subgame perfect investment strategy depends on time discounting.

\textbf{Organization of the paper}: The remainder of this paper is organized
as follows. In Section $2$, we describe the model and formulate the
objective. Section $3$ introduces the notion of subgame perfect strategies. Section $4$ introduces the value function. Section $5$ presents
the main results. Section $6$ deals with the utility weighted discount rate and Section 7 presents numerical results. We then wrap up our findings in the Conclusion. The Appendix contains various proofs, and the description of the numerical scheme.

\section{The Model}

\subsection{The Financial Market}
Consider a financial market consisting of a savings account and one stock
(the risky asset). The inclusion of more risky assets can be achieved by
notational changes.  We assume a benchmark deterministic time horizon $T$. 
The stock price per share follows an exponential Brownian motion 
\begin{equation}\label{eq121_1}
dS(t)=S(t)\left[ \mu(t,S(t)) \,dt+\sigma(t,S(t)) \,dW(t)\right] ,\quad 0\leq t\leq T,
\end{equation}%
where $\{W(t)\}_{t\geq 0}$ is a $1-$dimensional Brownian motion on a
filtered probability space, $(\Omega ,\{\mathcal{F}_{t}\}_{0\leq t\leq T},\mathbb{P}).$  The filtration $\{\mathcal{F}_{t}\}_{0\leq t\leq T}$ is the completed
filtration generated by $\{W(t)\}_{0\leq t\leq T}.$ The savings account accrues interest at the riskless rate $r(t,S(t)),$ for some function $r(t,z).$

 Let us denote \be \label{eq121_2}
 \theta \triangleq \frac{\mu
-r}{\sigma},\ee
 \emph{the market price of risk}.

We place ourselves in a Markovian setting.
The stock mean rate of return $\mu$ and volatility $\sigma$ will be functions
of the running time $t$ and the stock price $S(t)$.
We make the following assumption on $r, \theta, \sigma$:

 \begin{assumption}[Standing Assumption]\label{ass121_1}
 Suppose that the functions
 \begin{enumerate}
 \item  $\sigma, r, \theta,  z\frac{\partial r}{\partial z},  z\frac{\partial \sigma}{\partial z}, z\frac{\partial \theta}{\partial z}$ are bounded in $[0, T]\times (0, \infty)$ and Lipschitz continuous.
 \item  There exists a positive constant $\sigma_0$ such that $\sigma \geq \sigma_0$.
\end{enumerate}
 \end{assumption}

In our model there is one agent who is continuously investing in the stock, 
the money market, and consuming. At every time $t$, the agent chooses $\pi(t)$, the ratio of wealth invested in the
risky asset and  $c(t)$ the ratio of wealth consumed. Given an adapted process $\{\pi (t),c(t)\}_{0\leq t\leq T}$, the
equation describing the dynamics of wealth ${X^{\pi ,c}(t)}$ is given by : 
\begin{eqnarray}\label{eq121_3}
dX^{\pi ,c}(t) &=&[r(t, S(t))+\pi(t)\sigma(t, S(t) )\theta(t, S(t))-c(t)]X^{\pi ,c}(t)dt+\pi
(t)\sigma(t,S(t) )X^{\pi,c}(t)dW(t)   \\
X^{\pi ,c}(0) &=&x_0, \nonumber
\end{eqnarray}%
the initial wealth $x_0$ being exogenously specified.

\subsection{Time Preferences and Risk Preferences }
The time preference reflects the economic agent's preference for immediate utility over delayed utility. We now define discount functions and discount rates.
\begin{definition}
A discount function $h:D=\left\{ (t,s)|0\leq t\leq s\leq T\right\}\rightarrow \sR$ is a $C^1$, positive function satisfying $h(t,t)=1$.
\end{definition}
For a discount function $h$, we define the backward discount rate as 
\begin{equation}\label{eq121_5}
\rho_b(t,s) = \frac{\partial h(t,s)}{\partial t}\times \frac{1}{h(t,s)}.
\end{equation}
The forward discount rate is 
\begin{equation}\label{eq121_6}
\rho_f(t,s) = \rho_h(t,s) = - \frac{\partial h(t,s)}{\partial s}\times \frac{1}{h(t,s)}.
\end{equation}

In the case $h(t,s)=H(s-t)$ for a certain $C^1$ function $H$ on $[0,T]$, we get: $\rho_f(t,s)=\rho_b(t,s)=-\frac{H'(s-t)}{H(s-t)}$. If it is obvious from the context, we just write $\rho(t,s)$ and call it the discount rate.
For any continuous function $y :[0,T]\times (0, \infty) \rightarrow (0,\infty)$ , denote $|| y || = \sup_{(t,s)\in [0,T]\times (0, \infty) } |y(t,s)|$ the norm sup of $y$.

\begin{remark}
We take a discount form to be of the general form $h(t,s)$ because in some problems such as the one studied in \cite{Ek_Mb_Pi1}, one has to account for stochastic time horizons $T$ ( e.g. the time of death of the agent).
In that case, after some transformations, the discount function will take the general form $h(t,s)$.
One can normalize dividing $h(t,s)$ by $h(t,t).$
\end{remark}

We define next the agent's \emph{risk preferences}.
The agent gets satisfaction $U(C)$ from consuming an amount $C$. We assume $U$ belongs to the class of constant relative risk aversion (CRRA), power type utilities:
\be 
U(x) = U_{\g}(x) = \frac{x^{\g}}{\g}, \g<1,\, \g\neq 0,\, x>0.
\ee

\subsection{The Expected Utility}
 Let us now define the admissible strategies.

\begin{definition}
 \label{def121_1} An $\mathbb{R}^{2}$-valued
stochastic process $\{\pi (t),c(t)\}_{0\leq t\leq T}$ is called an
admissible strategy process if: \begin{enumerate}
\item it is progressively measurable with respect to filtration $\{\mathcal{F}_{t}\}_{0\leq t\leq T},$
\item it satisfies \begin{equation}\label{eq12_3}
c(t)>0\,\, \text{ for all $0\leq t\leq T$ almost surely and } {X}^{\pi ,c}(T)>0,\text{ almost surely }
\end{equation} 
\item and is required that 
\begin{equation}  \label{eq121_4}
\mathbb{E} \sup_{0\leq s\leq T} |c(s)X^{\pi,c}(s)|^{\gamma}<\infty \; \; , \; \;
 \mathbb{E}
\sup_{0\leq s\leq T} |{X}^{\pi,c}(s)|^{\gamma}<\infty, \, \text{ a.s.%
}
\end{equation}
\end{enumerate}
\end{definition}

The last set of inequalities are purely technical and are satisfied for e.g. bounded strategies.

In order to evaluate the performance of an investment-consumption
strategy the agent uses the expected utility criterion. For an
admissible strategy process $\{{\pi }(s),c(s)\}_{s\geq 0}$ and its
corresponding wealth process $\{X^{\pi ,c}(s)\}_{s\geq 0}$, we denote the expected
(intertemporal) utility  by

\begin{eqnarray}  \label{eq122_1}
J(t, x,\pi ,c)&=& \mathbb{E}_{t} \bigg[ \int_{t}^{T} h(t,s)
U_{\g}(c(s)X^{\pi,c}(s))\,ds + h(t,T) {U}_{\g}%
(X^{\pi,c}(T))\bigg].
\end{eqnarray}
Here $\sE_t$ denotes the conditional expectation given time $t,$ and $X^{\pi,c}(t)=x.$ 
A natural objective for the decision maker is to maximize the above expected
utility criterion.

\begin{definition}\label{def14_2}
For an admissible strategy $(\pi,c)$ and
$(t,z,x)\mapsto v(t,z,x)$ a $C^{1,2,2}$ function, define the operator $\mathcal{A}^{\pi,c}$ by:
 \begin{equation}\label{eq14_5}
 \mathcal{A}^{\pi,c}v(t,z,x)= z\mu\frac{\partial v}{\partial z}(t,z,x)+(r+\sigma\theta \pi -c)x
\frac{\partial v}{\partial x}(t,z,x)+ \frac{\sigma^2z^2}{2}%
\frac{\partial ^{2}v}{\partial z^{2}}(t,z,x)$$$$+\frac{1}{2}\sigma ^{2}x^2\pi ^{2}%
\frac{\partial ^{2}v}{\partial x^{2}}(t,z,x)+\sigma^2 z \pi x\frac{\partial^2 v}{\partial z\partial x}(t,z,x).
\end{equation}
\end{definition}

Let $X^{\pi,c}$ be the corresponding wealth process, then by It\^{o}'s lemma, \eqref{eq14_5} yields
$$\frac{\partial v}{\partial t}(t,z,x)+\mathcal{A}^{\pi,c}v(t,z,x)= \frac{d}{dt}\sE_t[dv(t,S(t),X^{\pi,c}(t))],$$ given $S(t)=z.$ 
Thus, $\mathcal{A}^{\pi,c}v(t,z,x)$ measures the average variation of $v$ when an infinitesimal time $dt$ passes and the agent follows the strategy $(\pi,c)$.

Let $\hat{V}(t_0; t,z,x)$ satisfy the HJB equation
\begin{eqnarray}  \label{eq122_3}
&&\frac{\partial \hat{V}}{\partial t}+\!\!\!\sup_{(\pi ,c)\; \textrm{admissible}}\bigg[ \mathcal{A}^{\pi,c}\hat{V}+U_{\g} (xc(t))\bigg] -\rho_h(t_0,t) \hat{V}(t_0; t,z,x) = 0, \nonumber \\
\end{eqnarray}
for $t_0\leq t\leq T$ and $\hat{V}(t_0; T,z,x)=U_{\g} (x).$ Then, given that $S(t)=z,$ the  quantity $\hat{V}(t_0,t_0,z,x)$ is the optimal value function starting at time $t_0,$ i.e.,
\begin{eqnarray}  \label{eq122_2}
\hat{V}(t_0; t_0,z,x)  = \sup_{(\pi,c) \; \textrm{admissible}} J(t_0, x,\pi ,c).
\end{eqnarray}
The optimal strategy $(\hat{\pi}, \hat{c})$ is the one realizing the sup in \eqref{eq122_2}. However, because the discount function $h$  is not exponential, the discount rate
 $\rho_h(t_0,t)=-\frac{\frac{\partial h(t_0,t)}{\partial s}}{h(t_0,t)}$ is not constant.
  Therefore, the solution $\hat{V}$ of the HJB \eqref{eq122_3} depends on the starting point $t_0$ and so does the optimal strategy starting at $t_0$. As such, every time we change the starting point $t_0$,
  we get a different optimal strategy. Time inconsistency will bite, that is, a strategy that will be considered optimal at time $0$ will not be considered so at later times, so it will
not be implemented unless the agent at time $t_0$ can constrain his successive selves to follow the time $t_0$- optimal strategy (the precommitment strategy) at all times $t_0\leq t\leq T.$ 

The agent could implement two types of strategies: he could \emph{precommit} at time $t_0$ to follow the optimal strategy and stay with it until time $T;$
or he could implement a time consistent strategy  that takes into account the fact that his decisions will change in the future. This is the object of the next section.

\section{Subgame Perfect Strategies}

Following \cite{Ek_Pi1} and \cite{Ek_Mb_Pi1} we introduce a special class of time consistent strategies called \emph{subgame perfect strategies} .
Assume that the agent at time $t$ can commit his successors up to time $t+\epsilon 
$, with $\epsilon \rightarrow 0,$ then he seeks for strategies which are
optimal to implement right now conditioned on them being implemented in the
future. This is made precise in the following formal definition.

\begin{definition}\label{def13_1}
An admissible trading strategy $\{\bar{\pi}(s),\bar{c}(s)\}_{0\leq s\leq T}$ is a \emph{subgame perfect strategy} if there exists a map $%
G=(G_{\pi}, G_{c}):[0,T]\times ( 0,\infty ) \times (0,\infty ) \rightarrow \mathbb{R}\times
(0,\infty )$ such that for any $t\in [0,T] , x >0$ 
\begin{equation}\label{eq13_1}
{\lim \inf_{\epsilon \downarrow 0}}\frac{J(t,x,\bar{\pi},\bar{c}%
)-J(t,x,\pi _{\epsilon },c_{\epsilon})}{\epsilon }\geq 0,
\end{equation}%
where:%
\begin{equation}\label{eq13_2}
\bar{\pi}(s)={G_{\pi}(s,S(s),\bar{X}(s))},\quad \bar{c}(s)=G_{c}(s,S(s),\bar{X}%
(s)),
\end{equation}%
and the wealth process $\bar{X}(s)= X^{\bar{\pi},\bar{c}}(s) $ is a solution of
the stochastic differential equation (SDE): 
\begin{eqnarray}\label{eq13_3}
d\bar{X}(s)&=&\bar{X}(s)[r(s, S(s))+G_{\pi}(s,S(s),\bar{X}(s))\sigma(s,S(s))\theta(s, S(s)) -G_{c}(s,S(s),\bar{X}(s))]ds \nonumber \\
&&+G_{\pi}(s,S(s),\bar{X}(s)\sigma(s, S(s)) )\bar{X}(s) dW(s).
\end{eqnarray}

The process $\{{\pi }_{\epsilon }(s),{c}_{\epsilon
}(s)\}_{s\in \lbrack t,T]}$ mentioned above is another
investment-consumption strategy, with corresponding wealth process $\{X_{\epsilon}(s)\}_{s\in \lbrack t,T]},$ defined by 
\begin{equation}\label{eq13_4}
\pi _{\epsilon }(s)=%
\begin{cases}
G_{\pi}(s, S(s), X_{\epsilon}(s)),\quad s\in \lbrack t,T]\backslash E_{\epsilon ,t} \\ 
\pi (s),\quad s\in E_{\epsilon ,t},%
\end{cases}
\end{equation}%
\begin{equation}\label{eq13_5}
c_{\epsilon }(s)=%
\begin{cases}
G_c(s,S(s),X_{\epsilon}(s)),\quad s\in \lbrack t,T]\backslash E_{\epsilon ,t} \\ 
c(s),\quad s\in E_{\epsilon ,t}%
\end{cases}
\end{equation}%
Here $E_{\epsilon,t}=[t,t+\epsilon],$ and $\{{\pi}(s),{c}(s)\}_{s\in E_{\epsilon,t} }$ is any strategy for which $\{{\pi}%
_{\epsilon}(s),{c}_{\epsilon} (s)\}_{s\in[t,T]}$ is an
admissible strategy starting at $X_{\epsilon}(0)=x.$

\end{definition}

In other words, time consistent strategies are Markov strategies that penalize
deviations on infinitesimally small time intervals. 

\section{The Value Function}
The paper \cite{Ek_Mb_Pi1} uses the value function methodology to characterize subgame perfect strategies. In our setting the value function can be written as a function $V(t,z,x)$ of time $t$, stock price $z$ and wealth $x.$  We start with a definition.

\begin{definition}\label{def14_1}
 Let $V : [0,T] \times (0,\infty)\times  (0,\infty) \rightarrow \mathbb{R},$ be a $C^{1,2,2}$ function that is concave in the $x$ variable.
Suppose that $\{\bar{\pi }(s),\bar{c}(s)\}_{s\in [0,T]}$ is an admissible strategy,

\begin{equation}\label{eq14_1}
\bar{\pi}(s)={G_{\pi}(s,S(s),\bar{X}(s))}\;\;,\quad \bar{c}(s)=G_{c}(s,S(s),\bar{X}%
(s)), 
\end{equation}%
given by
\begin{equation}\label{eq14_2}
\!\!\!\!\! G_{\pi}(t,z,x)=-\frac{\theta(t,z) \frac{\partial V}{\partial x}(t,z,x)+z\sigma(t,z)\frac{\partial^2 V}{\partial z\partial x}(t,z,x)}{x \sigma(t,z)
\frac{\partial ^{2}V}{\partial x^{2}}(t,z,x)}\;\;,\ \ G_{c}(t,z,x)=\frac{1}{x}\left( \frac{%
\partial V}{\partial x}(t,z,x)\right)^{\frac{1}{\g-1}}, \end{equation}%
and $\bar{X}(s)$ is the wealth process given by:
\begin{eqnarray}\label{eq14_3}
d\bar{X}(s)&=&[r(s)+\sigma(s)\theta(s) G_{\pi}(s,S(s),\bar{X}(s))-G_{c}(s,S(s),\bar{X}(s))]\bar{X}(s)ds \nonumber\\
&&+\sigma(s) G_{\pi}(s,S(s),\bar{X}(s))\bar{X}(s)dW(s). 
\end{eqnarray}

We shall say that $V$ is a value function if for all $(t, z, x)\in [0,T]\times (0,\infty)\times(0,\infty),$ 
\begin{equation}\label{eq14_4}
V(t,z,x)=J(t,x,\bar{\pi},\bar{c}), 
\end{equation}
given that $S(t)=z.$

\end{definition}

The economic interpretation is very natural: if one implements [\eqref{eq14_1}, \eqref{eq14_2}, \eqref{eq14_3}] and computes the
corresponding value of the agent's criterion starting from $S(t)=z,\; X(t)=x$ at time $t,$ one gets precisely $V\left( t,z,x\right).$  
In the next section, we give the main results of this paper.

\section{Main Results}

\subsection{The Extended HJB}

\begin{definition}
Let $V : [0,T] \times (0,\infty)\times  (0,\infty) \rightarrow \mathbb{R},$ be a $C^{1,2,2}$ function that is concave in the $x$ variable, and $(\bar{\pi}, \bar{c})$ an admissible strategy.
Then $(V, \bar{\pi}, \bar{c})$ solves the extended HJB if
 \begin{eqnarray}  \label{eq151_1}
&&\frac{\partial V}{\partial t}(t,z,x)+\sup_{(\pi ,c)\; \textrm{admissible}}\big\{ \mathcal{A}^{\pi,c}V(t,z,x)+U_{\g} (xc(t))\big\} \nonumber\\
&=&\sE_t \left[\int_t^T \frac{\partial h(t,s)}{\partial t}U_{\g}(\bar{c}(s)X^{\bar{\pi},\bar{c}}(s) ) ds + \frac{\partial h(t,T)}{\partial t}U_{\g}(X^{\bar{\pi},\bar{c}}(T) )\right],
\end{eqnarray}
along with the boundary condition $V(T,z,x)= U_{\g}(x),$ 
and $(\bar{\pi},\bar{c})$ satisfies
\be  \label{eeq151_2}
(\bar{\pi},\bar{c}) = \mbox{argmax}_{(\pi ,c)\; \textrm{admissible}} \{ \mathcal{A}^{\pi,c}V(t,z,x)+U_{\g} (xc(t))    \} 
\ee
\end{definition}

\begin{theorem}\label{thm151_1}
 Let $(V, \bar{\pi}, \bar{c})$ solve the extended HJB. Then $(\bar{\pi}, \bar{c})$ is given by \eqref{eq14_1}, \eqref{eq14_2},  $V$ is a value function, and $(\bar{\pi}, \bar{c})$ is a subgame perfect strategy.
\end{theorem}



The proof is given in the Appendix. Next, we define a strategy dependent discount rate that we call utility weighted discount rate.

\subsection{The Utility Weighted Discount Rate}
\begin{definition}\label{def14_3}
Let $(\pi,c)$ be an admissible strategy. The utility weighted discount rate corresponding to $(\pi,c)$ is defined as the process
\begin{equation}\label{eq14_6}
Q^{\pi,c}(t)= \frac{\sE_t \int_t^{T}\frac{\partial h(t,s)}{\partial t} U_{\g}(c(s)X^{\pi,c}(s))ds+\frac{\partial h(t,T)}{\partial t} U_{\g}(X^{\pi,c}(T)) }{\sE_t \int_t^{T} h(t,s) U_{\g}(c(s)X^{\pi,c}(s))ds+h(t,T)U_{\g}(X^{\pi,c}(T))}.
\end{equation}
If $(\pi,c)=(\bar{\pi},\bar{c})$ then $Q^{\bar{\pi},\bar{c}}(t)$ is called the subgame perfect discount rate.


\end{definition}

The right hand side of the extended HJB \eqref{eq151_1} is $$Q^{\bar{\pi},\bar{c}}(t) V(t,z,x).$$
In the exponential discounting case $h(t,s)=\exp(-\rho(s-t))$, the quantity $Q^{\pi, c}(t)$ simplifies to 
$$Q^{\pi, c}(t)=\frac{\sE_t[\int_t^T \rho e^{-\rho(s-t)}U(c(s)X^{\pi,c}(s))ds+\rho e^{-\rho(T-t)}U(X^{\pi,c}(T) )]]}{\sE_t[\int_t^T e^{-\rho(s-t)}U(c(s)X^{\pi,c}(s))ds+e^{-\rho(T-t)}U(X^{\pi,c}(T) )]}=\rho.$$
In general, $Q^{{\pi},{c}}(t)$ behaves like an average discount rate.

Our goal is to compute $Q^{\bar{\pi},\bar{c}}(t).$ It turns out that $$Q^{\bar{\pi},\bar{c}}(t)=\mathbb{Q}(t,S(t), X^{\bar{\pi},\bar{c}}(t)),$$ and 
is $\mathbb{Q}$ is the fixed point of an operator as outlined in the subsection below.

Knowing $\mathbb{Q}$ leads to knowing the subgame perfect strategy $(\bar{\pi},\bar{c})$. Indeed, the extended HJB \eqref{eq151_1} 
 becomes a classical HJB 
 \begin{eqnarray}\label{eq152_1}
 \frac{\partial V}{\partial t}+\sup_{(\pi ,c)\; \textrm{admissible}}\!\!\!\!\!  \big\{ \mathcal{A}^{\pi,c}V(t,z,x)+U_{\g} (xc(t))\big\} = \mathbb{Q}(t,z,x)V(t,z,x), \label{eq152_1b}
 \end{eqnarray}
that can be tackled through standard techniques. Having found $V$, we use equations \eqref{eq14_1}, and \eqref{eq14_2} to obtain $(\bar{\pi},\bar{c})$.

\subsection{Fixed Point Characterization of Subgame Perfect Discount Rate}
Let a function $\mathbb{\hat{Q}}(t,z,x)$ be given, and $\hat{V}(t,z,x)$ the solution of HJB \eqref{eq152_1b} with right hand side $$\mathbb{\hat{Q}}(t,z,x) \hat{V}(t,z,x).$$ Let  $(\hat{\pi},\hat{c})$ given as in \eqref{eq14_1}, \eqref{eq14_2}, $X^{\hat{\pi},\hat{c}}(s)$ be the corresponding wealth process, and $Q^{\hat{\pi},\hat{c}}(t)$ as in \eqref{eq14_6}. Define the operator $L$ by 
$$ L[\mathbb{\hat{Q}}] (t,z,x)= Q^{\hat{\pi},\hat{c}}(t).$$
Then, the subgame perfect discount rate $\mathbb{Q}(t,z,x)$ is a fixed point of operator $L.$

The financial economic intuition of this result is clear, as the subgame perfect strategies are in theory fixed points of an intrapersonal game. The novelty
of our approach is to devise a concrete fixed point iteration that solves the extended HJB equation. Let us point out that this fixed point characterization 
can be applied for general utilities. ln the following we exploit the special structure of power type utilities which makes the subgame perfect discount rate $\mathbb{Q}(t,z,x)$ independent of $x,$
hence $$\mathbb{Q}(t,z,x)=\mathbb{Q}(t,z).$$ Moreover the fixed point characterization of $\mathbb{Q}$ gets a simplified form which we outline below. Define $p$  the inverse of the relative risk aversion:
\begin{equation}\label{eq152_2}
p=\frac{1}{1-\g}.
\end{equation}
In what follows, unless we specify otherwise, we will write
 \be  \label{eq152_3}
 \rho=\rho_b : (t,s) \in D \rightarrow \sR, (t,s)\mapsto \frac{\frac{\partial h(t,s)}{\partial t}}{h(t,s)}.
 \ee
 
 We work under the assumption
 $$ ||\rho||<\infty, $$
 and this condition is fairly general as it covers typical discount functions such as exponential, hyperbolic, and generalized hyperbolic.
 Next, we introduce the following operator.

We define a space of functions in which we want to find the fixed point $\mathbb{Q}(t,z)$.
In order to simplify the calculations, we work with $y=\log S$ and $Y(t) = \log S(t)$.
The process $Y$ satisfies the SDE
\be
dY(s) = ( \tilde{\mu}(s,Y(s)) - \frac{\tilde{\sigma}^2}{2}(s,Y(s)) )ds +\tilde{\sigma}(s,Y(s)) dW(s),\,\, Y(t)=y,
\ee
where 
$$\tilde{\mu}(t,y) = \mu(t,e^y), \tilde{\sigma}(t,y)=\sigma(t,e^y), \tilde{r}(t,y) = r(t,e^y).$$ 
The solution to this SDE will sometimes be denoted $Y^{t,y}(s).$ 
For a fixed $\lambda >0 $, and $\phi$ a bounded continuous function of $(t,y)\in [0,T]\times \sR$,   define
$$ ||\phi||_{0,\lambda}=\sup_{ (t,y)\in [0,T]\times \sR} e^{\lambda t} |\phi(t,y)|,$$
$$ ||\phi||_{L,\lambda}=\sup_{ t\in [0,T], y_1, y_2 \in \sR, y_1 \neq y_2} e^{\lambda t} \frac{|\phi(t,y_2)-\phi(t,y_1)|}{|y_2-y_1|}.$$

Notice that $||.||_{0,\lambda}$ is equivalent to the sup norm and $||.||_{L,\lambda}$ is equivalent to the Lipschitz semi-norm. 
Let $\delta$ be a positive scalar, and
$$\mathbb{B}_{\delta,\lambda}=\{\phi\in C([0,T]\times \sR),  ||\phi|| \leq ||\rho||, ||\phi ||_{L,\lambda} \leq \delta \}.$$
Then $\mathbb{B}_{\delta, \lambda}$ is a complete set under the norm $||.||_{0,\lambda}$. For $\phi$ an element of $\mathbb{B}_{\delta,\lambda},$ define the operators:
\begin{eqnarray}\label{eq152_5}
&&\tilde{F}_1[\phi](t,y)=\sE \bigg[ \int_t^{T} \frac{\partial h(t,s)}{\partial t} \exp\left(p\g \int_t^s (\tilde{r}+\frac{\tilde{\theta}^2}{2}-\phi) (u, Y(u)) du+p\g\int_t^s \tilde{\theta}(u, Y(u)) dW(u) \right)ds \nonumber\\
&&+\frac{\partial h(t,T)}{\partial t} \exp\left(p\g\int_t^T (\tilde{r}+\frac{\tilde{\theta}^2}{2}-\phi)(u, Y(u))du+p\g \int_t^T \tilde{\theta}(u, Y(u)) dW(u) \right)\bigg|  Y(t)=y \bigg],
 \end{eqnarray}
\begin{eqnarray}\label{eq152_6}
&&\tilde{F}_0[\phi](t,y)=\sE \bigg[ \int_t^{T} h(t,s) \exp\left(p\g \int_t^s (\tilde{r}+\frac{\tilde{\theta}^2}{2}-\phi)(u, Y(u)) du+p\g\int_t^s \tilde{\theta}(u, Y(u)) dW(u) \right)ds \nonumber\\
&&+ h(t,T) \exp\left(p\g \int_t^T (\tilde{r}+\frac{\tilde{\theta}^2}{2}-\phi)(u, Y(u))du+p\g \int_t^T \tilde{\theta}(u, Y(u)) dW(u) \right)\bigg|  Y(t)=y \bigg],
 \end{eqnarray}
 and 
 \begin{equation}\label{eq152_7}
 \tilde{F}[\phi](t,y)=\frac{\tilde{F}_1[\phi](t,y)}{\tilde{F}_0[\phi](t,y)}.
 \end{equation}

 \begin{theorem}\label{thm152_2} 
The operator $\tilde{F}$ has a unique fixed point, denoted $\mathbb{\tilde{Q}},$ on $\mathbb{B}_{\delta,\lambda}.$
\end{theorem}
The proof is given in the Appendix.

For $\phi$ an element of $C([0,T]\times (0,\infty)),$ define the operators:
\begin{eqnarray}\label{eq152_5}
&&F_1[\phi](t,z)=\sE \bigg[ \int_t^{T} \frac{\partial h(t,s)}{\partial t} \exp\left(p\g \int_t^s (r+\frac{\theta^2}{2}-\phi) (u, S(u))du+p\g\int_t^s \theta (u, S(u)) dW(u) \right)ds \nonumber\\
&&+\frac{\partial h(t,T)}{\partial t} \exp\left(p\g\int_t^T (r+\frac{\theta^2}{2}-\phi)(u, S(u))du+p\g \int_t^T \theta (u, S(u)) dW(u) \right) \bigg | S(t) =z  \bigg],
 \end{eqnarray}
\begin{eqnarray}\label{eq152_6}
&&F_0[\phi](t,z)=\sE \bigg[ \int_t^{T} h(t,s) \exp\left(p\g \int_t^s (r+\frac{\theta^2}{2}-\phi)(u, S(u))du+p\g\int_t^s \theta(u, S(u)) dW(u) \right)ds \nonumber\\
&&+ h(t,T) \exp\left(p\g \int_t^T (r+\frac{\theta^2}{2}-\phi)(u,S(u))du+p\g \int_t^T \theta(u, S(u)) dW(u) \right) \bigg | S(t) =z  \bigg],
 \end{eqnarray}
 and 
 \begin{equation}\label{eq152_7}
 F[\phi](t,z)=\frac{F_1[\phi](t,z)}{F_0[\phi](t,z)}.
 \end{equation}

In light of Theorem \ref{thm152_2}, $\mathbb{Q}$ defined by

\be\label{fp}
 \mathbb{Q}(t,z)= \mathbb{\tilde{Q}}(t,\log z),\ee
is a fixed point of $F.$ Having obtained $\mathbb{Q}$, we can characterize $V$ through a linear PDE. This is the object of the next subsection.

\subsection{Characterization of The Value Function}

\begin{proposition}\label{prop153_1}
There exists a unique $C^{1,2}([0,T]\times (0,\infty))$ solution $v$ to the linear parabolic PDE :
\begin{eqnarray}\label{eq153_1}
0 &=& \frac{\partial v}{\partial t}(t,z)+\frac{\sigma^2 z^2}{2}\frac{\partial^2 v}{\partial z^2}(t,z)+z(r+p\sigma\theta)(t,z)\frac{\partial v}{\partial z}(t,z) \nonumber \\
&&+p(\g r+\frac{\g p\theta^2}{2}-\mathbb{Q})(t,z)v(t,z) +1\\
v(T,z)&=&1. \nonumber
\end{eqnarray} 
Here $\mathbb{Q}$ is the fixed point of $F$ defined by \eqref{fp}. Moreover, $v$ is bounded away from zero, and $z\frac{\partial v}{\partial z}$ is bounded.
\end{proposition}
The proof is given in the Appendix.

We now present the main result of this paper.

\begin{theorem}\label{thm153_1}
Let $v$ be the solution of \eqref{eq153_1}, and $V$ the function
$$V(t,z,x)=v(t,z)^{1-\g}\frac{x^{\g}}{\g}.$$
Define the strategies by
\be \label{eq153_6} 
\bar{c}(t)= \bar{G}_{c} (t, S(t)), \quad \quad \bar{\pi}(t)=\bar{G}_{\pi} (t, S(t)),  
\ee\be \label{eq153_7} 
\bar{G}_{c} (t,z)=\frac{1}{v(t,z)},\quad  \quad \bar{G}_{\pi} (t,z)=\frac{\theta(t,z)}{(1-\gamma)\sigma(t,z)}+\frac{z\frac{\partial v}{\partial z} (t,z)}{v(t,z)}.  
\ee
Then $(V, \bar{\pi}, \bar{c})$ solves the extended HJB. Consequently $(\bar{\pi}, \bar{c})$ is a subgame perfect strategy.

\end{theorem}

The proof is given in the Appendix.

\begin{remark}
This result did not mention the uniqueness of subgame perfect strategies. It turns out that the extended HJB has a uniques solution in the class of wealth independent strategies.
That is when $ {c}(t)= {G}_{c} (t, S(t)),  {\pi}(t)={G}_{\pi} (t, S(t)),$ then it can be shown that  $V(t,z,x)=v(t,z)^{1-\g}\frac{x^{\g}}{\g},$ and $v$ solves \eqref{eq153_1}.

Let us notice that our methodology also applies to the settings of \cite{Ek_Pi1} and \cite{Ek_Mb_Pi1}, providing a fixed point characterization
of the subgame perfect strategies. In that paradigm, the parabolic equation which characterizes the value function becomes a first order
ordinary differential equation with a non local term. Our result yields the uniqueness of  the subgame perfect strategies of 
 \cite{Ek_Pi1} and \cite{Ek_Mb_Pi1}  in the class of wealth independent strategies and it provides a fixed point iterative scheme to compute them. 
\end{remark}

\section{Further Interpretation of Subgame Perfect Discount Rate  }

The following Corollary to Theorem \ref{thm153_1} summarizes our findings about the utility weighted discount rate and the subgame perfect strategy.

\begin{corollary}\label{cor26_1}
The subgame perfect strategy is the precommitment strategy of an agent with discount rate $\mathbb{Q}(t,S(t))$ on $t \leq T.$ 
\end{corollary}
This is also the finding of  \cite{Bj_Kh_Mu1}. Next, we can obtain bounds for $\mathbb{Q}(t,z).$ 

Let  $$h(t,s)=f(s-t) \textrm{ for } s\geq t, $$
and $\alpha(t,s,z),$ $\rho(t,s)$ denote the quantities
\begin{eqnarray*}
&&\alpha(t,s,z)=\sE \bigg[ \exp\left(p\g \int_t^s (r+\frac{\theta^2}{2}-y)(u, S(u))du+p\g\int_t^s \theta (u, S(u)) dW(u) \right)  \bigg| S(t)=z \bigg] \\
&& \rho(t,s) = -\frac{\frac{\partial h(t,s)}{\partial s}}{h(t,s)} =-\frac{f'(s-t)}{f(s-t)}=R(s-t).
 \end{eqnarray*}
 Then $\mathbb{Q}(t,z)$ can be written as
\begin{eqnarray*}
&&\mathbb{Q}(t,z)=\frac{ \int_t^{T} R(s-t)f(s-t) \alpha(t,s,z) ds + R(T-t)f(T-t)\alpha(t,T,z) }{\int_t^{T} f(s-t) \alpha(t,s,z) ds + f(T-t)\alpha(t,T,z)}. \\
 \end{eqnarray*}
Therefore
 \begin{equation}\label{eq16_1}
\inf_{x\in [0,T-t]} R(x) \leq \mathbb{Q}(t,z) \leq \sup_{x\in [0,T-t]} R(x).
\end{equation}

If the agent follows the precommitment strategy starting at $t=0,$ then he uses  at time $s$ the following discount rate $\rho(0,s)=-\frac{f'(s)}{f(s)}=R(s).$
%
In the case of the generalized hyperbolic discounting, $f(x)=(1+a x)^{-\frac{b}{a}}. \exp(-\rho x)$ 
with positive constants $a$ and $b$ and non negative constant $\rho,$  \be R(x)=-\frac{f'(x)}{f(x)}=\rho+\frac{b}{1+a x}\ee is decreasing in $x$.
Thus \be
R(T-t) \leq \mathbb{Q}(t,z) \leq R(0).
\ee

The following graphs show the bounds for the utility weighted discount rate $\mathbb{Q}$:

\begin{figure}[htbp]
\vspace*{-30mm}
\begin{center}
\includegraphics[scale=0.46]{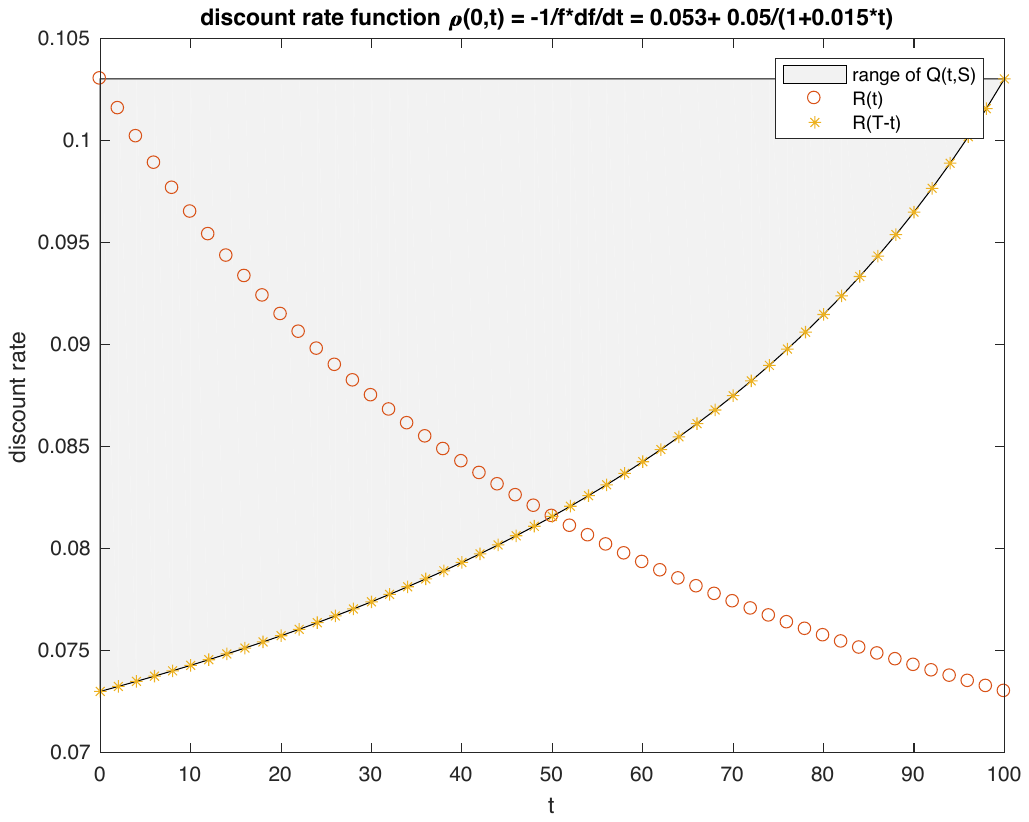}
\vspace*{-30mm}
\caption{The range of $\mathbb{Q}(t,S) $ for hyperbolic discounting is given by the shaded area. The decreasing function is the discount rate $R(t)=\rho(0,t).$  }
\label{range_of_Q }
\end{center}
\end{figure}

Next we present the linear parabolic PDEs  of the precommitment strategy.
An admissible strategy $(\hat{\pi}, \hat{c})$ is called time 0- optimal if \be
J(0,x,\hat{\pi}, \hat{c}) = \sup_{ (\pi,c)\,\, \textrm{admissible} } J(0,x, {\pi}, {c} ). \ee
In that case, the value function $\hat{V}$ satisfies $$\hat{V}(0,S,x)= J(0,x,\hat{\pi}, \hat{c}),$$ given that $S(0)=S.$ As noted before, $(\hat{\pi}, \hat{c})$ is called precommitment strategy.

The subgame perfect strategy $(\bar{\pi}, \bar{c})$ has the corresponding value function 
\be
{V}(t,z,x)= J(t,x,\bar{\pi}, \bar{c}),\ee
when $S(t)=z.$ We will see in the following result that $V$ and $\hat{V}$ are characterized by the same equations but with different discount rates.
\begin{proposition}\label{prop17_1}
For all $(t,z,x )\in [0,T]\times (0,\infty)\times (0,\infty)  $ : \be \hat{V}(t,z,x) = \hat{v}(t,z)^{1-\g} U_{\g}(x) \textrm{ and } V(t,z,x) = {v}(t,z)^{1-\g} U_{\g}(x). \ee
 The functions $\hat{v}$ and ${v}$ satisfy the linear parabolic PDEs:
 \begin{equation}\label{eq17_2}
\frac{\partial \hat{v}}{\partial t}(t,z)+\frac{\sigma^2 z^2}{2} \frac{\partial^2\hat{v}}{\partial z^2}(t,z)+(r+p\sigma\theta)z \frac{\partial\hat{v}}{\partial z}(t,z)  +p\left[\g r+\frac{\g p\theta^2}{2}-\rho(0,t) \right]\hat{v}(t,z)
+1 = 0,  
\end{equation}
\begin{equation}\label{eq17_1}
\frac{\partial v}{\partial t}(t,z)+\frac{\sigma^2 z^2}{2} \frac{\partial^2{v} }{\partial z^2}(t,z)+(r+p\sigma\theta)z\frac{\partial{v}}{\partial z}(t,z) +p\left[\g r+\frac{\g p\theta^2}{2}-\mathbb{Q}(t,z)\right]v(t,z)
+1 = 0,  
\end{equation}
with final condition $\hat{v}(T,z) = v (T,z)=1.$

%
%
%
 The precommitment strategy $(\hat{\pi},\hat{c})$ and subgame perfect strategy $(\bar{\pi},\bar{c})$ are given by:
 \begin{eqnarray}\label{eq17_4}
&& \hat{c}(t,z)=\frac{1}{\hat{v}(t,z)} \quad , \quad \bar{c}(t,z)=\frac{1}{v(t,z)}\\
&& \hat{\pi}(t,z)=\frac{\theta(t,z)}{(1-\gamma)\sigma(t,z)}+\frac{p z \frac{\partial\hat{v}}{\partial z}  (t,z) }{\hat{v}(t,z)} \quad , \quad  \bar{\pi}(t,z)=\frac{\theta(t,z)}{(1-\gamma)\sigma(t,z)}+\frac{p z\frac{\partial{v} }{\partial z}(t,z) }{v(t,z)}.
 \end{eqnarray}
 \end{proposition}
The proof is straightforward and thus omitted. We can see that the parabolic PDEs of $\hat{v}$ and $v$ differ only through the discount rate term.
The precommitment agent discounts at the rate $\rho(0,t)$ while the subgame perfect agent discounts at the rate $\mathbb{Q}(t,S(t))$.


\subsection{The Investment Strategy and Discounting}
It turns out that the precommitment investment strategy  does not depend on the discount rate. Indeed, this finding is made formal in the following Proposition. 

\begin{proposition}\label{IND}
 The precommitment investment strategy $\hat{\pi}$ is independent of the discount rate $\rho(0,t).$
\end{proposition}

The proof is given in the Appendix.

The subgame perfect investment strategy, on the other hand, depends on the discount rate and this dependence makes it observationally different when compared to the precommitment investment strategy.

\section{ Numerical Analysis}\label{NA}

The subgame perfect discount rate $\mathbb{Q}$ is defined as the fixed point of the non linear operator $F$. One can compute it numerically, as shown in the Numerical Scheme subsection of the Appendix. Then we use it to calculate $v(t,z)$ by Monte Carlo simulation, in light of Feyman-Kac representation.

We consider 2 cases for the parameters of our model.
\begin{enumerate}
\item {Constant volatility model.}
\; It is a model where all the coefficients are assumed to be constants: $\sigma(s)=\sigma, \theta(s)=\theta, r(s)=r$.
This model was previously considered in \cite{Ek_Mb_Pi1}. 
\item {Constant elasticity of variance model (CEV).}
\end{enumerate}
The CEV model is a model where the instantaneous volatility is specified to be a power function of
the underlying spot price $\sigma(z)=\alpha z^{\beta},$ $\alpha>0$ is the volatility scale parameter, and
$\beta$ is the elasticity parameter of the local volatility: $\beta=\frac{1}{\sigma(z)}\frac{\partial \sigma}{\partial z}.$ If $\beta=0$, one retrieves the Black Scholes Merton model, and
for $\beta=-\frac{1}{2},$ one retrieves the square root model of Cox and Ross.

In the remainder, we let $\beta <0$. As explained in \cite{Li_Me1}, 
the spot volatility is a decreasing function of the asset price.
The stock price volatility increases as the
stock price declines. This shows the leverage effect in equity markets.
The implied volatility in this model turns out to have a skewed shape. That is what makes this model attractive in the finance world.
However, \cite{De_Sh1} shows that there always exists arbitrage in such markets.

Note that the volatility could go to infinity when the stock price goes to zero. 
We want to avoid those anomalies since we are not concerned with defaults, and as such  a minimum $\sigma_m$ and a maximum $\sigma_M$ values are set for the volatility $\sigma$.
Let us assume that the volatility at time 0 equals $\sigma_0,$ and the stock price is $S_0.$
Next, choose $\alpha = \sigma_0 S_0^{-\beta}$ and one can write 
$\sigma(z)=\sigma_0 (\frac{z}{S_0})^{\beta}$. Moreover, $\sigma(z)$ is set to $\sigma_M$ if $\sigma(z)\geq \sigma_M,$ and $\sigma(S)$ is set to $\sigma_m$ when $\sigma(z)\leq \sigma_m$. 
 The parameters presented in the table are used in our numerical experiments.

\begin{center}
\begin{tabular}{ c| c c c c  c c c c c} 
 \hline
 Parameters & $r(t,z)$ & & $\beta$   & $\sigma$  & $\theta(t,z)$ & $\sigma_m$  &  $\sigma_M$   \\ 
 \hline
 Values         &         0.05  &   &  -0.4 &              $0.3 (\frac{z}{10})^{\beta}   $      & $  6 \sigma (t,z)$ &         0.15                   & 0.45    \\ 
 \hline
\end{tabular}
\end{center}

We take $\g = -5$ and $$h(t,s)=H(s-t)=(1+\alpha_1(s-t))^{-\frac{\beta_1}{\alpha_1}}\exp(-\rho_1(s-t))$$ with $\alpha_1 = 1.0 , \beta_1 = 0.02 , \rho_1=0.02 $. 
The superscript "PC" represents the precommitment optimal strategies while "TC" represents time consistent (subgame perfect) strategies.

For the constant volatility model, we have chosen the market parameters $(r,\theta,\sigma) = (0.05, 0.2777, 0.30)$.

\subsection{Investment Strategy}\label{IS}

Let us turn now to subgame perfect and precommitment investment strategies. As it was already shown in Proposition \ref{IND} the precommitment investment
strategy does not depend on the discount rate. This is also the case for the subgame perfect investment strategy when stock volatility is constant in which case
the two investment strategies are equal. Figures 4 and 5 exhibit subgame perfect and precommitment investment strategies and their differences for constant versus
non constant stock volatility; for the later the difference does not have a fixed sign.

\begin{figure}
\centering
\vspace*{-30mm}
\includegraphics[scale=0.60]{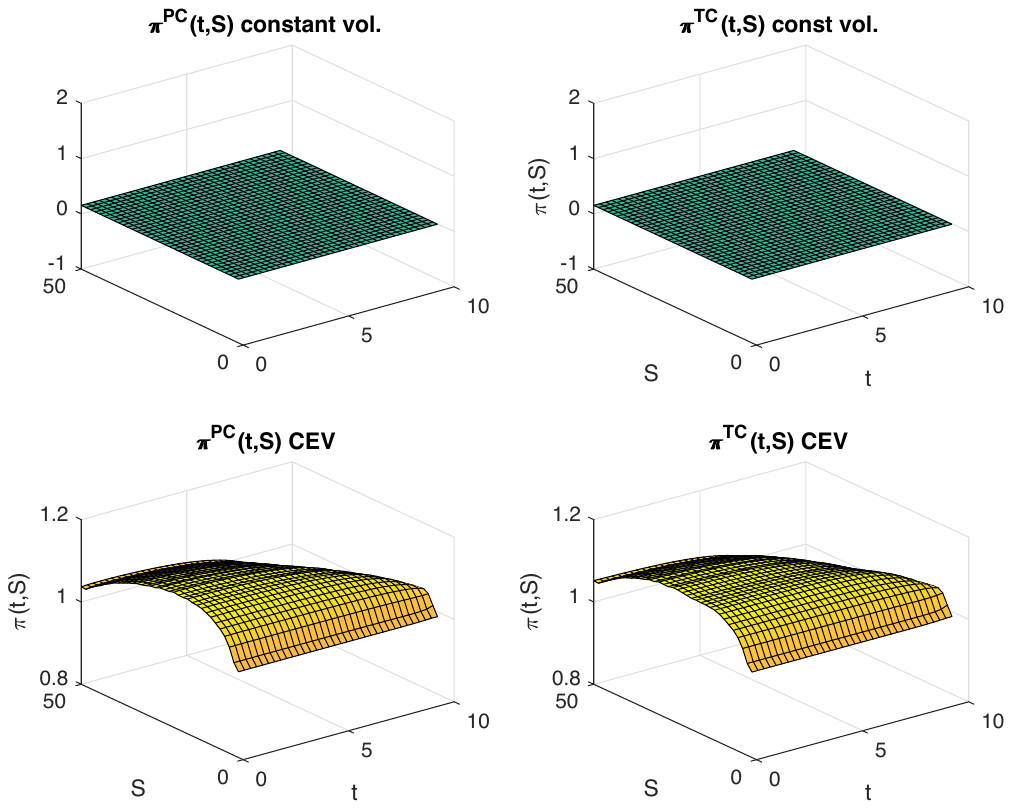}
\vspace*{-40mm}
        \caption{ Comparing $\pi$ for different volatilities and strategies : $\pi^{PC}=\hat{\pi} = \bar{\pi}=\pi^{TC}$ when the parameters $r, \theta, \sigma$ are all constants.
        Theoretically, we get $\bar{\pi}=\hat{\pi}=\frac{\theta}{\sigma(1-\g)}$ is independent of $z$}
        \label{fig:pi }
    \end{figure}

\begin{figure}
\centering
\vspace{-30mm}
    \begin{subfigure}[b]{0.45\textwidth}
        \centering
       \includegraphics[width=1.35\textwidth]{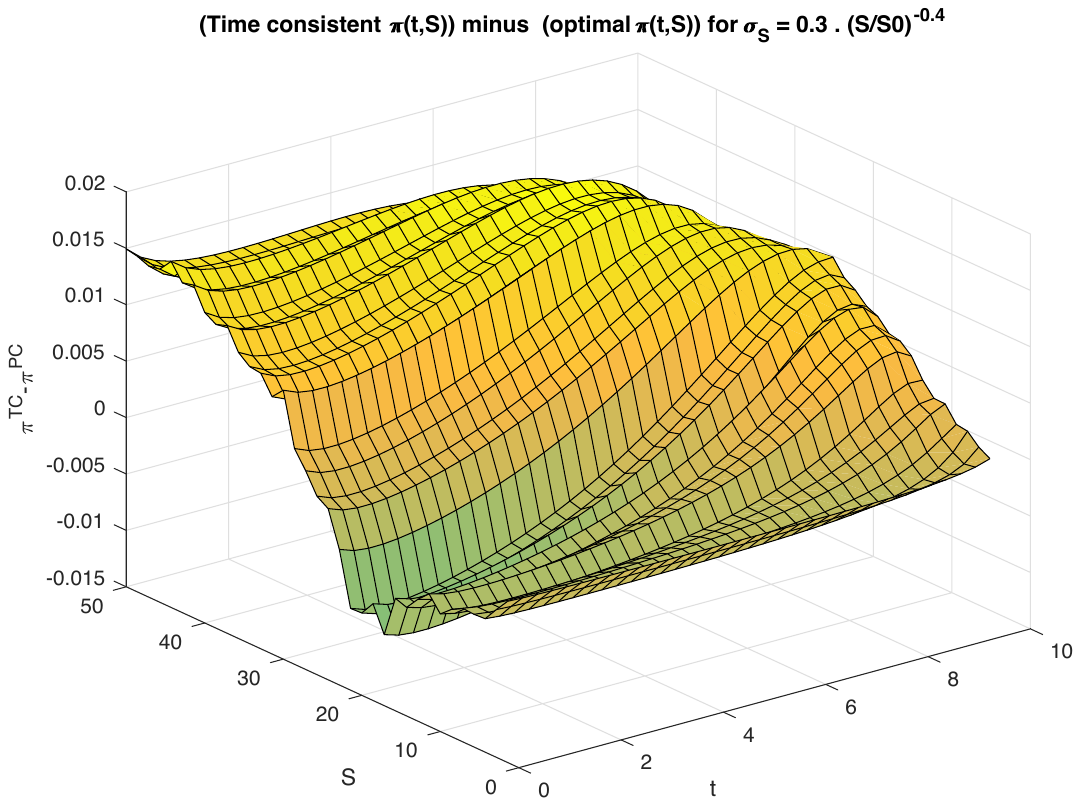}
                \vspace*{-20mm}
        \caption{ $\bar{\pi}(t,z)-\hat{\pi}(t,z)$ in the case of CEV. The difference does not have a fixed sign but changes with the stock price $S$ }
      \label{fig:pi CEV}
    \end{subfigure}
    \hfill
     \begin{subfigure}[b]{0.45\textwidth}
\centering
      \includegraphics[width=1.35\textwidth]{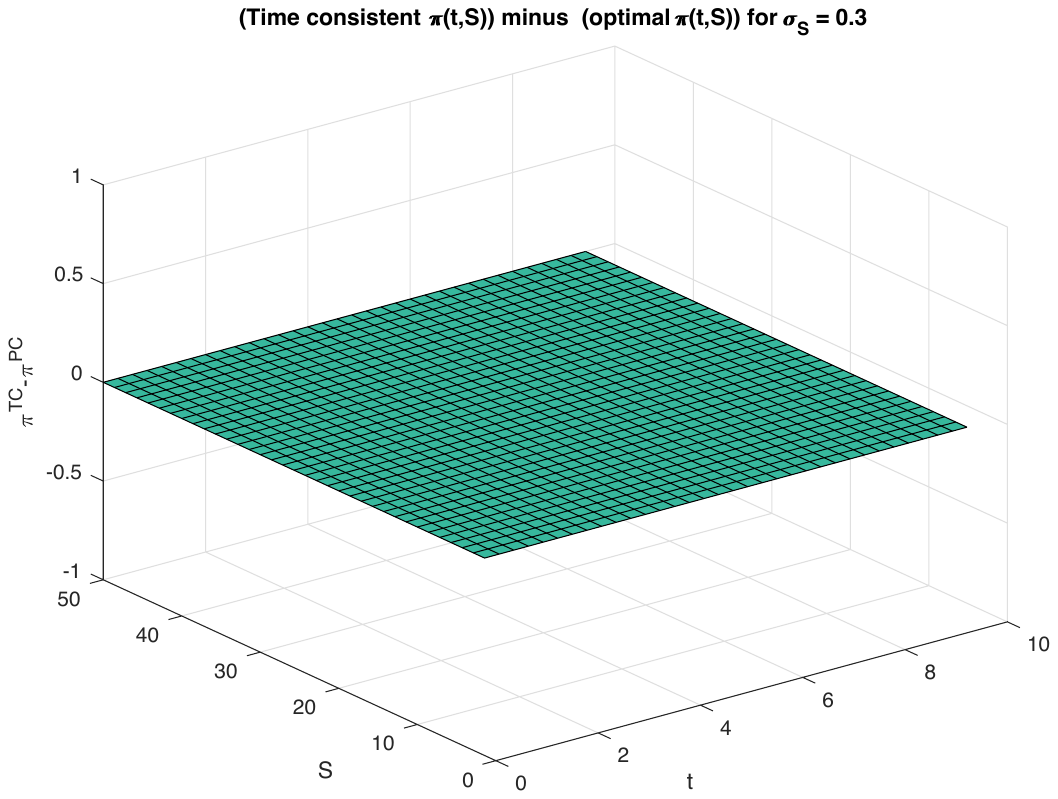}
                        \vspace*{-20mm}
       \caption{$\bar{\pi}(t,z)-\hat{\pi}(t,z)$ for constant volatility. Actually, $\bar{\pi}=\hat{\pi}$ if $r, \theta, \sigma$ are all constant. }
        \label{fig:pi constant volatility}
\end{subfigure}
            \vspace*{10mm}
    \caption{Study of $\hat{\pi}(t,z), \bar{\pi}(t,z)$ for $\g=-5$}
    \label{fig:Study of pitS for negative gamma }
\end{figure}

\subsection{Consumption}

\begin{figure}
\centering
\vspace*{-25mm}
    \begin{subfigure}[b]{0.45\textwidth}
        \centering
        \vspace*{-50mm}
       \includegraphics[width=1.36\textwidth]{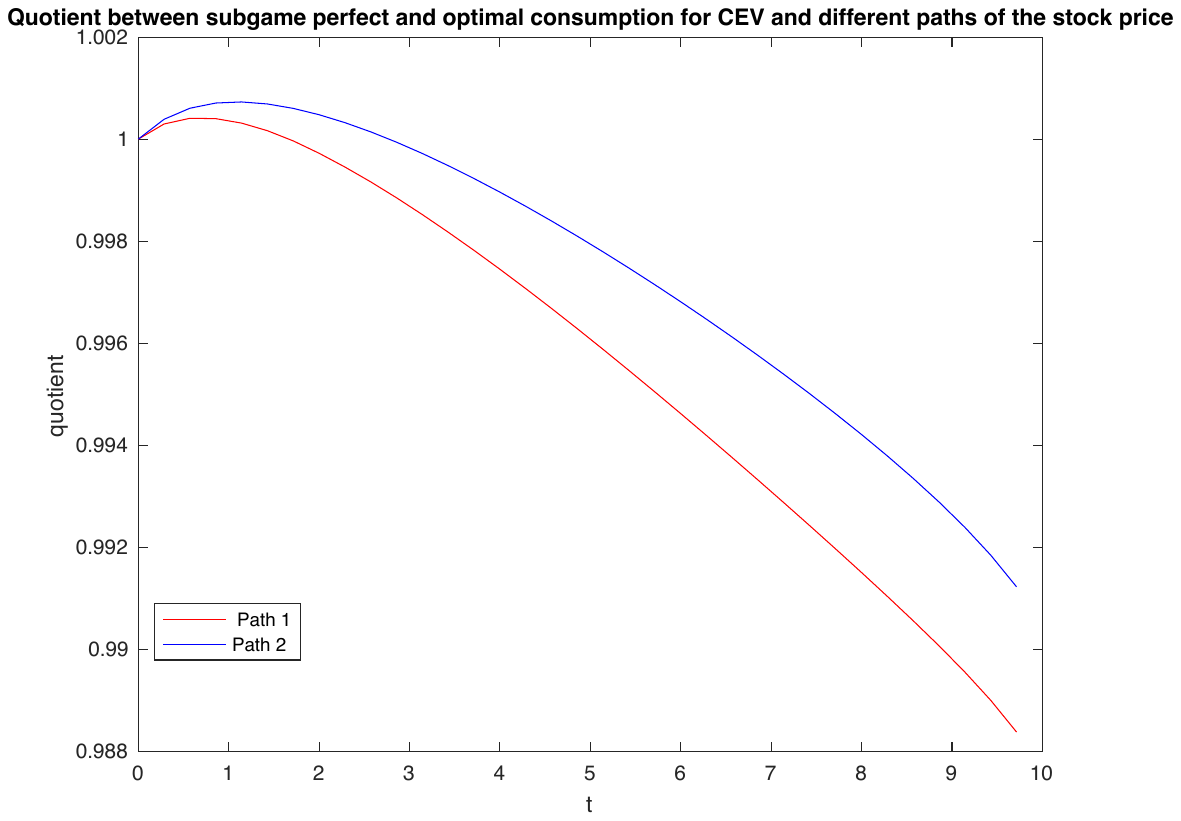}
            \vspace*{-30mm}
        \caption{ Graph of the consumption quotients for CEV type of volatility}
        \label{fig:quotientc CEV}
   \end{subfigure}
    \hfill
     \begin{subfigure}[b]{0.45\textwidth}
        \centering
        \includegraphics[width=1.36\textwidth]{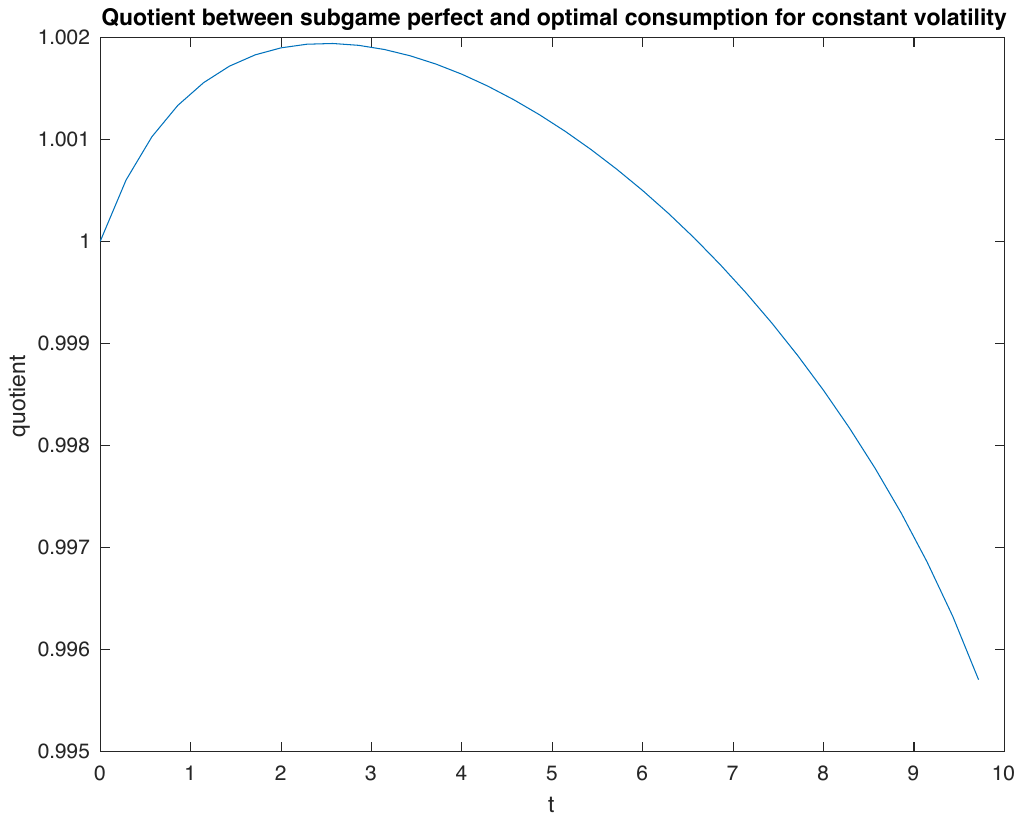}
            \vspace*{-30mm}
        \caption{Graph of the consumption quotients for constant volatility}
        \label{fig:quotientc constant volatility}
  \end{subfigure}
    \vspace*{5mm}
    \caption{Study of $\frac{\bar{c}(s)\bar{X}(s)}{  \hat{c}(s)\hat{X}(s)}$ for $\g=-5$}
    \label{fig:Study of quotients of consumption.}
\end{figure}

We see from the plots in Figure 3, that when compared to the agent that is following a precommitment strategy, the subgame perfect agent has:
a higher consumption in the short term, and a lower consumption in the medium to long term. This fact is explained by: 1) for small $t, R(t)\geq \mathbb{Q}(t,s),$ 2) for big $t, R(t)\leq \mathbb{Q}(t,s)$ and
3) the higher the discount rate the lower the consumption conform \cite{Ek_Pi1}.

\section{Conclusion}

This work considers the Merton portfolio management problem when the volatility of the stock is stochastic, and the time discount rate
 is non-constant. The optimal strategies are time inconsistent and referred to as precommitment strategies since they are  implementable only through a commitment mechanism.
 The subgame perfect strategies on the other hand are time consistent . The later are characterized through a value function. A new approach of finding the subgame perfect strategies through fixed point iterations is established and implemented. The stock's volatility makes it more difficult to find the value function, and the novel approach is to consider the utility weighted discount rate; this quantity, strategy dependent, turns out to be the fixed point of a certain operator. This finding allows us to disentangle the search for the value function as the solution of a parabolic linear equation. The precommitment  investment strategy turns out to be independent of the discount rate while the subgame perfect investment strategy depends on the discounting; thus, the two strategies are observationally different. However, in constant drift/volatility models, precommitment  and subgame perfect investment strategies coincide and as such they are independent of the discount rate.


\section{Appendix}
\markboth{}{}

\proof (Theorem \ref{thm151_1})
Let $V\in C^{1,2,2},$ concave in $x$, satisfying \eqref{eq151_1} and $(\bar{\pi}, \bar{c})$ satisfying \eqref{eeq151_2}.
The fact that $(\bar{\pi}, \bar{c})$ is given by \eqref{eq14_1}, \eqref{eq14_2} comes straight from the first order conditions.
We want to show that $V$ is a value function and $(\bar{\pi},\bar{c})$ is a subgame perfect strategy.
First, we have to show that  $V(t,z,x)=J(t,x,\bar{\pi},\bar{c}),$ when $S(t)=z.$ As before, $\bar{X}$ represents the process $X^{\bar{\pi}, \bar{c}}$.
Dynkin's theorem states that the process

$$V(s,S(s),\bar{X}(s))-\int_0^s \big(\frac{\partial V}{\partial t}(u,S(u),\bar{X}(u))+\mathcal{A}^{\bar{\pi},\bar{c}}V(u,S(u),\bar{X}(u))\big) du \textrm{ is a martingale. } $$
 Therefore
\begin{equation}\label{eeq01_1}
\sE_t[V(T,S(T),\bar{X}(T))]=V(t,S(t),x)+\sE_t \bigg[\int_t^T (\frac{\partial V}{\partial t}(u,S(u),\bar{X}(u))+\mathcal{A}^{\bar{\pi},\bar{c}}V(u,S(u),\bar{X}(u)) du\bigg],
\end{equation}

Let the function $\delta$ be defined by 
\be
\delta(t,s,z,x)=\begin{cases}
 \sE_t [U_{\g}(\bar{c}(s)\bar{X}(s))| S(t)=S] & \text{if $t\leq s<T$}\\ 
 \sE_t [U_{\g}(\bar{X}(T))| S(t)=S]& \text{if $ s=T$}.
\end{cases}
\ee

By using \eqref{eq151_1} in the RHS of expression \eqref{eeq01_1}, we get:

\begin{eqnarray}\label{eeq01_2}
&&RHS=V(t,S(t),x)+\sE_t \bigg[\int_t^T \frac{\partial V}{\partial t}(s,S(s),\bar{X}(s))+\mathcal{A}^{\bar{\pi},\bar{c}}V(s,S(s),\bar{X}(s)) ds\bigg] \nonumber \\
&&=V(t,S(t),x)+\sE_t \big[\int_t^T -U_{\g}( \bar{c}(s)\bar{X}(s))+\sE_s\big[ \int_s^T \frac{\partial h}{\partial s}(s,u)U_{\g}(\bar{c}(u)\bar{X}(u))du \nonumber \\
&&+\frac{\partial h}{\partial s}(s,T)U_{\g}(\bar{X}(T))\big]ds\big]\\
&& =V(t,S(t),x)+\int_t^T(-\delta(t,s,S(t),x)+\int_s^T \frac{\partial h}{\partial s}(s,u)\delta(t,u,S(t),x)du
+\frac{\partial h}{\partial s}(s,T)\delta(t,T,S(t),x)\big)ds.\nonumber \\
\end{eqnarray}
The last equality comes from the law of iterated conditional expectations.
We then use the relation:
\begin{eqnarray*}\label{eq01_3}
&&\frac{\partial}{\partial s}\big(\int_s^T h(s,u)\delta(t,u,z,x)du+h(s,T)\delta(t,T,z,x)\big)\\
&&=-\delta(t,s,z,x)+\int_s^T \frac{\partial h}{\partial s}(s,u)\delta(t,u,z,x)du+\frac{\partial h}{\partial s}(s,T)\delta(t,T,z,x),
\end{eqnarray*}

so that 
\begin{eqnarray*}\label{eq01_4}
&&V(t,z,x)+\int_t^T \frac{\partial}{\partial s}\big(\int_s^T h(s,u)\delta(t,u,z,x)du+h(s,T)\delta(t,T,z,x)\big)ds\\
&=&V(t,z,x)+h(T,T)\delta(t,T,z,x)-\int_t^T h(t,u)\delta(t,u,z,x)du-h(t,T)\delta(t,T,z,x)\\
&&=V(t,z,x)+U_{\g}(x)-\int_t^T h(t,u)\delta(t,u,z,x)du-h(t,T)\delta(t,T,z,x).
\end{eqnarray*}
Since $$\sE_t[V(T,S(T),\bar{X}(T))]=\delta(t,T,S(t),x), \quad  $$
then it follows that
\begin{eqnarray}\label{eq01_5}
V(t,z,x) &=& \int_t^T\!\!\! h(t,u)\delta(t,u,z,x)du+h(t,T)\delta(t,T,z,x)=J(t,x,\bar{\pi},\bar{c}),
\end{eqnarray}
when $S(t)=z.$ The next step is to show that $(\bar{\pi},\bar{c})$ is a subgame perfect strategy. For this, we study the $\lim\inf$ of the quotient $$\frac{J(t,x,\bar{\pi},\bar{c})-J(t,x,\pi_{\epsilon},c_{\epsilon})}{\epsilon}.$$

\begin{eqnarray}\label{eq01_6}
\lefteqn{\frac{J(t,x,\bar{\pi},\bar{c})-J(t,x,\pi_{\epsilon},c_{\epsilon})}{\epsilon} } \nonumber\\
&=& \frac{1}{\epsilon}\sE_{t}\bigg[\int_t^{T}h(t,u) \left(U_{\g} (\bar{X}(u)\bar{c}(u)) - U_{\g} (X_{\epsilon}(u)c_{\epsilon}(u))\right)du \bigg] \\
&+&\frac{1}{\epsilon}\sE_{t}[ h(t,T) (U_{\g} (\bar{X}(T)) - U_{\g} (X_{\epsilon}(T)))], \nonumber 
\end{eqnarray}

\be
 \frac{J(t,x,\bar{\pi},\bar{c})-J(t,x,\pi_{\epsilon},c_{\epsilon})}{\epsilon} = K_1(\epsilon)+K_2(\epsilon)+K_3(\epsilon),
\ee
where the $K_1, K_2, K_3$ are obtained by introducing intermediate terms in the expression above.

\begin{eqnarray*}\label{eq01_7}
K_1(\epsilon)&=& \frac{1}{\epsilon}\sE_{t}\left[\int_t^{t+\epsilon}h(t,u)\left(U_{\g} (\bar{X}(u)\bar{c}(u)) - U_{\g} (X_{\epsilon}(u)c_{\epsilon}(u))\right)du \right]\\
K_2(\epsilon)&=&\frac{1}{\epsilon}\sE_{t}\left[\int_{t+\epsilon}^{T}(h(t+\epsilon,u)-h(t,u))(U_{\g}(X_{\epsilon}(u)c_{\epsilon}(u))- U_{\g} (\bar{X}(u)\bar{c}(u))du\right]\\
&+&  \frac{1}{\epsilon}\sE_{t}\bigg[(h(t+\epsilon,T)-h(t,T))(-U_{\g}(X_{\epsilon}(T))+ U_{\g}(\bar{X}(T)))\bigg]\\
K_3(\epsilon)&=&\frac{1}{\epsilon}\sE_{t}\left[\int_{t+\epsilon}^{T}h(t+\epsilon,u)\left(U_{\g} (\bar{X}(u)\bar{c}(u)) - U_{\g} (X_{\epsilon}(u)c_{\epsilon}(u))\right)du\right]\\
&+&  \frac{1}{\epsilon}\sE_{t}\left[h(t,T)(U_{\g} (\bar{X}(T)) - U_{\g} (X_{\epsilon}(T)))\right].\\
\end{eqnarray*}
It is easy to see that \begin{equation}\label{eq01_8}
\lim_{\epsilon\rightarrow 0}K_1(\epsilon)=h(t,t)(U_{\g}(\bar{c}(t)x)-U_{\g}(c(t)x))=U_{\g}(\bar{c}(t)x)-U_{\g}(c(t)x).
\end{equation}
%
%
%
\begin{eqnarray}\label{eq01_9}
&&K_2(\epsilon)=\frac{1}{\epsilon}\int_{t+\epsilon}^{T}(h(t+\epsilon,u)-h(t,u))\sE_{t}\left[(U_{\g}(X_{\epsilon}(u)c_{\epsilon}(u))- U_{\g} (\bar{X}(u)\bar{c}(u))du\right] \nonumber\\
&+&\frac{1}{\epsilon}\sE_{t}\left[(h(t+\epsilon,T)-h(t,T))(U_{\g} (\bar{X}(T)) - U_{\g} (X_{\epsilon}(T)))\! \right]\\
&&=I_1(\epsilon)+I_2 (\epsilon), \nonumber
\end{eqnarray}
where $I_1=I_1(\epsilon)$ and $I_2=I_2(\epsilon)$ are given by:
\be  \label{eq01_9a}
I_1 = \frac{1}{\epsilon}\int_{t+\epsilon}^{T}\!\!\!(h(t+\epsilon,u)-h(t,u))
\sE_{t}\bigg[\bigg(\bigg( \frac{X_{\epsilon}(t+\epsilon)c_{\epsilon}(t+\epsilon)}{\bar{X}(t+\epsilon)\bar{c}(t+\epsilon)}\bigg)^{\g}-1\! \bigg) U_{\g} (\bar{c}(u)\bar{X}(u))du \bigg], 
\ee
\be \label{eq01_9b}
I_2= \frac{1}{\epsilon}(h(t+\epsilon,T)-h(t,T))\times
\sE_{t}\left[ \bigg(\left( \frac{X_{\epsilon}(t+\epsilon)}{\bar{X}(t+\epsilon)}\right)^{\g}-1\bigg)
U_{\g} (\bar{X}(T))\right]. 
\ee
We can calculate an upper bound for the integral term $I_1$:
\begin{eqnarray*}\label{eq01_10}
&&|I_1(\epsilon)| \leq \epsilon \int_{t+\epsilon}^{T}\bigg|\frac{\partial h(t_u^{\epsilon},u)}{\partial t}\bigg |\times
\sE_{t}\left[\left|\left( \frac{X_{\epsilon}(t+\epsilon)c_{\epsilon}(t+\epsilon)}{\bar{X}(t+\epsilon)\bar{c}(t+\epsilon)}\right)^{\g}-1\right ||U_{\g} (\bar{c}(u)\bar{X}(u))|\right]du,\\
\end{eqnarray*}
where $t_u^{\epsilon}\in[t,t+\epsilon]$ and by hypothesis
$g(t,u)=\sup_{t_0\in[t,t+1]} \big|\frac{\partial h(t_0,u)}{\partial t}\big|$ is integrable on $[t,T]$. Therefore
\begin{eqnarray*}\label{eq01_11}
&&|I_1(\epsilon)| \leq \epsilon \int_{t+\epsilon}^{T}g(t,u)\times
\sE_{t}\left[\left|\left( \frac{X_{\epsilon}(t+\epsilon)c_{\epsilon}(t+\epsilon)}{\bar{X}(t+\epsilon)\bar{c}(t+\epsilon)}\right)^{\g}-1\right ||U_{\g} (\bar{c}(u)\bar{X}(u))|\right]du. \\
\end{eqnarray*}
Since $X_{\epsilon}(t+\epsilon)\rightarrow x$ and $\bar{X}(t+\epsilon)\rightarrow x$, the integrand goes to 
$$g(t,u) \sE_{t}\!\left[\left|\left( \frac{c(t)}{\bar{c}(t)}\right)^{\g}-1\right ||U_{\g} (\bar{c}(u)\bar{X}(u))|\right], $$
 and by the dominated convergence Theorem, 
 \be \label{eq01_12}
 I_1(\epsilon)\rightarrow 0 \textrm{ when } \epsilon\rightarrow 0 . \ee
For the same reasons, $I_2 \rightarrow 0$ when $\epsilon \rightarrow 0$. Thus,
\begin{equation}\label{eq01_13}
K_2(\epsilon)\rightarrow 0 \textrm{ as } \epsilon\rightarrow 0.
\end{equation}
\begin{eqnarray*}\label{eq01_14}
K_3(\epsilon) &=& \frac{1}{\epsilon}\sE_t\big[V(t+\epsilon,S(t+\epsilon),\bar{X}(t+\epsilon))-V(t+\epsilon,S(t+\epsilon),X_{\epsilon}(t+\epsilon))\big]\\
&=&\frac{1}{\epsilon}\sE_t\big[V(t+\epsilon,S(t+\epsilon),\bar{X}(t+\epsilon))-V(t,S(t),x)\\
&&+V(t,S(t),x)-V(t+\epsilon,S(t+\epsilon),X_{\epsilon}(t+\epsilon))\big]\\
&=&\frac{1}{\epsilon}\sE_{t}\bigg[\int_t^{t+\epsilon}dV(u,S(u),\bar{X}(u))-\int_t^{t+\epsilon}dV(u,S(u),X_{\epsilon}(u)))\bigg]\\
&=&\frac{1}{\epsilon}\sE_{t}\bigg[\int_t^{t+\epsilon}(  \frac{\partial V}{\partial t}(u,S(u),\bar{X}(u))+\Ac^{\bar{\pi},\bar{c}}V(u,S(u),\bar{X}(u)))du\\ &-& \int_t^{t+\epsilon}( \frac{\partial V}{\partial t}(u,S(u), X_{\epsilon}(u)))  +\Ac^{\pi,c}V(u,S(u),X_{\epsilon}(u))))du\bigg],
\end{eqnarray*}
\begin{eqnarray}\label{eq01_15}
K_3(\epsilon) \rightarrow_{\epsilon\rightarrow 0} [ \frac{\partial V}{\partial t}(t,S(t),x)+\Ac^{\bar{\pi},\bar{c}}V(t,S(t),x)]-[\frac{\partial V}{\partial t}(t,S(t),x)+\Ac^{\pi,c}V(t,S(t),x)].
\end{eqnarray}
Combining the limits, given that $S(t)=z,$ we get
\begin{eqnarray}\label{eq01_16}
\lefteqn{\liminf_{\epsilon \rightarrow 0}\frac{J(t,x,\bar{\pi},\bar{c})-J(t,x,\pi_{\epsilon},c_{\epsilon})}{\epsilon} } \nonumber \\
&&= (U_{\g}(\bar{c}(t)x)+\Ac^{\bar{\pi},\bar{c}}V(t,z,x)) -(U_{\g}(c(t)x)+\Ac^{\pi,c}V(t,z,x)) \geq 0, 
\end{eqnarray}
in light of $(\bar{\pi}, \bar{c})$ satisfying \eqref{eq14_1}, \eqref{eq14_2}. This ends the proof.
\ep

 \proof (Theorem \ref{thm152_2}) In a first step we establish that

$$  \tilde{F}( \mathbb{B}_{\delta,\lambda})\subset \mathbb{B}_{\delta,\lambda}. $$

  Let us start with some estimates. Set
\begin{equation}
   a_{0}=\g p ( \tilde{r}+\frac{\tilde{\theta}^2}{2}), \quad \quad b_{0}=\g p\tilde{\theta}
\end{equation}

and

  \be 
Z^{\phi; t,y}(s)=\int_t^s (a_{0}(u, Y(u))-\g p {\phi}(u,Y(u)) ) du+\int_t^s b_{0}(u,Y(u))dW(u).
\ee
  
Then

  \be 
Z^{\phi; t,y}(s)\geq  e^{ -(p |\g| . ||\rho||+||{a}_0|| )(s-t) } e^{ \int_t^s b_{0}(u,Y(u))dW(u)}.
\ee

Moreover,

$$
 \sE_t e^{ \int_t^s b_{0}(u,Y(u))dW(u) }\geq \sE_t e^{ \left( \int_t^s b_{0}(u,Y(u))dW(u)-\int_t^s  \frac{1}{2}b^{2}_{0}(u,Y(u))du \right)}=1.
$$

 Therefore, one gets the following upper bound estimates  
\be
\sE_t[e^{Z^{\phi; t,y}(s)} ] \geq  e^{ -(p |\g| . ||\rho||+||{a}_0|| )(s-t) }= m_{1}(t,s)
\ee

For $t\in[0,T]$ and $y\in \sR$: 
\begin{eqnarray}\label{lb}
&&\tilde{F}_{0}[\phi](t,y)=\sE_t\bigg[ \int_t^T  h(t,s)e^{Z^{\phi; t,y}(s)}ds+h(t,T)  e^{Z^{\phi; t,y}(T) }\bigg] \nonumber\\
&&\geq \int_t^T  h(t,s)m_1(t,s)ds+h(t,T) m_1(t,T) = m(t)>m ,\nonumber\\
\end{eqnarray} 
 for some $m>0.$ 
 One has
 $$e^{k Z^{\phi; t,y}(s)} \leq  e^{ (|\g| p k ||\rho||+k ||{a}_0||)(s-t) } e^{ \int_t^s k b_{0}(u,Y(u))dW(u)}. $$
 
 Moreover
 
 $$e^{ \int_t^s k b_{0}(u,Y(u))dW(u)}\leq e^{ \frac{k^2}{2} ||{b}_0||^2 (s-t) } e^{ \left( \int_t^s k b_{0}(u,Y(u))dW(u)-\int_t^s \frac{k^2}{2} b^{2}_{0}(u,Y(u))du \right)},$$
 
 and
 
 $$\sE_t e^{ \left( \int_t^s kb_{0}(u,Y(u))dW(u)-\int_t^s  \frac{k^2}{2}b^{2}_{0}(u,Y(u))du \right)}=1. $$

 Therefore, we have the following upper bound estimates 
\be
\sE_t[e^{k Z^{\phi; t,y}(s)} ] \leq  e^{ (|\g| p k ||\rho||+k ||{a}_0|| +\frac{k^2}{2} ||{b}_0||^2 )(s-t) }= M_{k}(t,s),
\ee
where $k$ is a positive integer.

Let $\phi\in \mathbb{B}_{\delta, \lambda}.$ Since $\tilde{F}[\phi]$ is defined as a quotient of expectations that are continuous in $(t,y)\in [0,T]\times \sR,$ it follows that $\tilde{F}[\phi]$ is continuous. In light of $| \frac{\partial h(t,s)}{\partial t}|\leq  ||\rho||  |h(t,s)|,$ it follows that 
$$  ||\tilde{F}(\phi)|| \leq ||\rho||. $$

 
 Let $t\in [0,T]$ and $y_1, y_2 \in \sR,$ then

 \begin{eqnarray}\notag
\tilde{F}[\phi](t,y_2)-\tilde{F}[\phi](t,y_1) &=&\frac{\tilde{F}_{1}[\phi](t,y_2)}{\tilde{F}_{0}[\phi](t,y_2)}-\frac{\tilde{F}_{1}[\phi](t,y_1)}{\tilde{F}_{0}[\phi](t,y_1)}\\\label{eu}
&= &\frac{\tilde{F}_1[\phi] (t,y_2)-\tilde{F}_1[\phi](t,y_1)-\tilde{F}[\phi](t,y_1)(\tilde{F}_0[\phi] (t,y_2)-\tilde{F}_0[\phi](t,y_1))}{\tilde{F}_0[\phi](t,y_2)}.
\end{eqnarray}

Since $|\tilde{F}_{0}[\phi](t, y_2)| \geq m(t)>m$, we just need to find an upper bound for $|\tilde{F}_{0}[\phi](t,y_2)-\tilde{F}_{0}[\phi]|(t, y_1)|$.
Let
\be
\Pi^{\phi; t,y_1,y_2}_s = \sE_t (e^{Z_s^{\phi; t,y_2}}-e^{Z_s^{\phi; t,y_1}}).
\ee
In light of \eqref{eu}, 

$$  ||\tilde{F}(\phi)|| \leq ||\rho||, $$

\begin{eqnarray*}
|\tilde{F}_{0}[\phi](t,y_2)-\tilde{F}_{0}[\phi](t, y_1)| &\leq & \int_t^T  h(t,s)|\Pi^{\phi; t,y_1,y_2}_s|ds+h(t,T)  |\Pi^{\phi; t,y_1,y_2}_T|,
\end{eqnarray*}

and

\begin{eqnarray*}
|\tilde{F}_{1}[\phi](t,y_2)-\tilde{F}_{1}[\phi](t, y_1)| &\leq & ||\rho|| \int_t^T  h(t,s)|\Pi^{\phi; t,y_1,y_2}_s|ds+h(t,T)  |\Pi^{\phi; t,y_1,y_2}_T|,
\end{eqnarray*}

one gets
$$|\tilde{F}[\phi](t,y_2)-\tilde{F}[\phi](t,y_1)|\leq \frac{2||\rho||}{|\tilde{F}_{0}[\phi](t, y_2)|} \times  \int_t^T  h(t,s)|\Pi^{\phi; t,y_1,y_2}_s|ds+h(t,T)  |\Pi^{\phi; t,y_1,y_2}_T|. $$

By Cauchy Schwarz's inequality one gets:
\begin{eqnarray*}
|\Pi^{\phi; t,y_1,y_2}_s| \leq \sE_t |e^{Z_s^{\phi; t,y_2}}-e^{Z_s^{\phi; t,y_1}}| \leq \sE_t [e^{\max(Z_s^{\phi; t,y_2}, Z_s^{\phi; t,y_1}) } \times |Z_s^{\phi; t,y_2}-Z_s^{\phi; t,y_1}|] \\
\leq \sqrt{ M_2(t,s)} \sqrt{ \sE_t |Z_s^{\phi; t,y_2}-Z_s^{\phi; t,y_1}|^2}.
\end{eqnarray*}
In
\begin{eqnarray*}
 Z_s^{\phi; t,y_2}-Z_s^{\phi; t,y_1} =&& \int_t^s (a_0(u, Y_u^{t,y_2})-\g p \phi(u,Y_u^{t,y_2}) ) -(a_0(u, Y_u^{t,y_1})-\g p \phi(u,Y_u^{t,y_1}) ) du\\
 &&+\int_t^s b_0(u,Y_u^{t,y_2})-b_0(u,Y_u^{t,y_1})dW_u,
\end{eqnarray*}
the $du$ term is bounded by
$(||a_0||_{L,0}+|\g|p \delta e^{-\lambda u} ) |Y_u^{t,y_2}-Y_u^{t,y_1}|,$ and the $dW_u$ term is bounded by $||b_0||_{L,0} |Y_u^{t,y_2}-Y_u^{t,y_1}|.$ Thus, 
\begin{eqnarray*}
\sE_t[( Z_s^{\phi; t,y_2}-Z_s^{\phi; t,y_1} )^2]  & \leq & 2(s-t) \sE_t \int_t^s (||a_0||_{L,0}+|\g|p \delta e^{-\lambda u} )^2 |Y_u^{t,y_2}-Y_u^{t,y_1}|^2  du+2\int_t^s ||b_0||_L^2 . |Y_u^{t,y_2})-Y_u^{t,y_1}|^2 du \\
& \leq &\big( 4T(||a_0||_{L,0})^2+4T(|\g|p \delta)^2  \big(\frac{e^{-2\lambda t}-e^{-2\lambda s} }{2\lambda}\big) +2 ||b_0||_{L,0}^2 \big) \sE_t \int_t^s  |Y_u^{t,y_2}-Y_u^{t,y_1}|^2  du.
\end{eqnarray*}

We also have
\begin{eqnarray*}
 Y_s^{t,y_2}-Y_s^{ t,y_1} =&& y_2-y_1+ \int_t^s (\mu_0-\frac{\sigma_0^2}{2})(u, Y_u^{t,y_2}) - (\mu_0-\frac{\sigma_0^2}{2})(u, Y_u^{t,y_1})  du\\
 &&+\int_t^s \sigma_0(u,Y_u^{t,y_2})-\sigma_0(u,Y_u^{t,y_1})dW_u
\end{eqnarray*}

\begin{eqnarray*}
\sE_t [|Y_s^{t,y_2}-Y_s^{ t,y_1} |^2] \leq && 3(y_2-y_1)^2+ 3(s-t)(||\mu_0||_{L,0}+\frac{||\sigma_0^2||_{L,0}}{2})\sE_t \int_t^s ( Y_u^{t,y_2}-Y_u^{ t,y_1} )^2  du\\
 &&+3\sE_t \int_t^s (||\sigma_0||_{L,0})^2 (Y_u^{t,y_2}-Y_u^{t,y_1})^2 du
\end{eqnarray*}

And by Gronwall's inequality
\begin{eqnarray*}
\sE_t[( Y_s^{\phi; t,y_2}-Y_s^{\phi; t,y_1} )^2] \leq  3(y_2-y_1)^2e^{ \alpha(s-t)}
\end{eqnarray*}
where $\alpha = 3T(||\mu_0||_{L,0}+\frac{||\sigma_0^2||_{L,0}}{2})
 + 3( ||\sigma_0||_{L,0})^2 $.
 We get the following inequality
 
\be
\sE_t[( Z_s^{\phi; t,y_2}-Z_s^{\phi; t,y_1} )^2] \leq  \bigg( 12T(||a_0||_{L,0})^2+6T(|\g|p \delta)^2 \big(\frac{e^{-2\lambda t}-e^{-2\lambda s} }{\lambda}\big) +6 ||b_0||_{L,0}^2 \bigg) (y_2-y_1)^2 e^{ \alpha(s-t)} \times (s-t)
\ee 
 
Thus,

\begin{eqnarray*}
|\Pi^{\phi; t,y_1,y_2}_s| && \leq \sqrt{  M_2(t,s)} \sqrt{ \sE_t |Z^{\phi_2; t,y}(s)-Z^{\phi_1; t,y}(s)|^2} \\
&&\leq    \bigg(\sqrt{6}(||b_0||_{L,0}+\sqrt{2T}||a_0||_{L,0}) + \sqrt{6T}(|\g|p \delta) . \frac{e^{-\lambda t}}{\sqrt{\lambda}} \bigg)    |y_2-y_1| \sqrt{s-t}\times e^{(|\g| p ||\rho||+||a_0||+||b_0||^2/2+\alpha/2)(s-t)}. 
\end{eqnarray*} 
 
 Therefore, there is a constant $K>0$ independent of $t, y_1, y_2, \delta, \lambda$ such that
  \begin{eqnarray*}
|\tilde{F}[\phi](t,y_2)-\tilde{F}[\phi](t,y_1)| \leq K\big(1+    \frac{\delta e^{-\lambda t}}{\sqrt{\lambda}} \big)    |y_2-y_1|  \sqrt{T-t}
\end{eqnarray*}
The aim is to have $\forall t\in [0,T]$, 
$$ Ke^{\lambda t} \sqrt{T-t}(1+\frac{\delta e^{-\lambda t}}{\sqrt{\lambda}}) \leq \delta. $$
 
 One can choose $\lambda $ big enough, so that $K \frac{\sqrt{T}}{\sqrt{\lambda}} \leq \frac{1}{2},$
and if we choose $\delta$ big enough so that  $Ke^{\lambda T} \sqrt{T} \leq \frac{\delta}{2}$ leads to
 
 \begin{eqnarray*}
e^{\lambda t} |\tilde{F}[\phi](t,y_2)-\tilde{F}[\phi](t,y_1)| \leq\delta |y_2-y_1|.
\end{eqnarray*}

 Having established that

$$  \tilde{F}( \mathbb{B}_{\delta,\lambda})\subset \mathbb{B}_{\delta,\lambda},$$
 next we will show that the operator $\tilde{F}$ is a contraction when $t\in [0,T],$ and as such admits a fixed point. More precisely it
will be shown that

\be\label{contr}  || \tilde{F}[\phi_2]-\tilde{F}[\phi_1]||_{0,\lambda} \leq \frac{1}{2} ||\phi_2-\phi_1||_{0,\lambda}.
 \ee

The key quantity to analyze is
 
 \be
\Delta^{t,y}(s) = \sE_t[e^{Z^{\phi_2; t,y}(s)} - e^{Z^{\phi_1; t,y}(s)}].
\ee

By Cauchy Schwarz's inequality one gets,
\begin{eqnarray*}
\sE_t |e^{Z^{\phi_2; t,y}(s)}-e^{Z^{\phi_1; t,y}(s)}| \leq \sqrt{  M_2(t,s)} \sqrt{ \sE_t |Z^{\phi_2; t,y}(s)-Z^{\phi_1; t,y}(s)|^2}.
\end{eqnarray*}

Furthermore

\begin{eqnarray*}
| Z^{\phi_2; t,y}(s)-Z^{\phi_1; t,y}(s)| =&& \bigg |\int_t^s (a_0(u, Y^{t,y}(u))-\g p \phi_{2}(u,Y^{t,y}(u)) ) -(a_0(u, Y^{t,y}(u))-\g p \phi_{1}(u,Y^{t,y}(u)) ) du \bigg| \\
 \leq && |\g|   p  ||\phi_{2}-\phi_{1}||_{0,\lambda} \times \int_{t}^{s} e^{-\lambda u} du=\frac{ |\g|  p    }{\lambda}( e^{-\lambda t}-e^{-\lambda s}) ||\phi_{2}-\phi_{1}||_{0,\lambda} 
\end{eqnarray*}

Thus,

\begin{eqnarray}
&&|\Delta^{t,y}(s)| \leq \sE_t |e^{Z^{\phi_2; t,y}(s)}-e^{Z^{\phi_1; t,y}(s)}| \leq  \frac{  |\g|  p  \sqrt{  M_2(t,s)} }{\lambda} ||\phi_{2}-\phi_{1}||_{0,\lambda} \times | e^{-\lambda t}-e^{-\lambda s}|.
\end{eqnarray}

In light of 
 
\begin{equation}\label{eeu}
\tilde{F}[\phi_2](t,y)-\tilde{F}[\phi_1](t,y)=\frac{\tilde{F}_{1}[\phi_2](t,y)-\tilde{F}_{1}[\phi_1](t,y)}{\tilde{F}_{0}[\phi_2](t,y)}+\frac{\tilde{F}_{1}[\phi_1](t,y)(\tilde{F}_{0}[\phi_1]-\tilde{F}_{0}[\phi_2])}{\tilde{F}_{0}[\phi_1](t,y) \tilde{F}_{0}[\phi_2](t,y)},\end{equation}

$$  ||\tilde{F}(\phi_1)|| \leq ||\rho||, $$

$$
 | \tilde{F}_{0}[\phi_2](t,y)-\tilde{F}_{0}[\phi_1](t,y)| \leq   \int_t^T h(t,s)  |\Delta^{t,y}(s)| ds + h(t,T) |\Delta^{t,y}(T)|,
$$

$$
 | \tilde{F}_{1}[\phi_2](t,y)-\tilde{F}_{1}[\phi_1](t,y)| \leq  ||\rho|| \int_t^T h(t,s)  |\Delta^{t,y}(s)| ds + h(t,T) |\Delta^{t,y}(T)|,
$$

one gets the following upper bound

\begin{equation}
 e^{\lambda t} | \tilde{F}[\phi_2](t,y)-\tilde{F}[\phi_1](t,y)| \leq    \frac{2 ||\rho||}{ m} e^{\lambda t} \int_t^T h(t,s)  |\Delta^{t,y}(s)| ds + h(t,T) |\Delta^{t,y}(T)| \leq \frac{M}{\lambda} \times ||\phi_{2}-\phi_{1}||_{0,\lambda}
 \end{equation}
for some universal constant $M$ independent of $t$ and $\lambda.$ Next, one can choose $\lambda$ large enough so that

\be\label{ec1} \frac{M}{\lambda}\leq \frac{1}{2}
 \ee

 to yield 
 
 $$ || \tilde{F}[\phi_2]-\tilde{F}[\phi_1]||_{0,\lambda} \leq \frac{1}{2} ||\phi_2-\phi_1||_{0,\lambda}.
 $$
 
  Thus, the operator $\tilde{F}$ is a contraction and this finishes the proof.

\ep

\proof (Proposition \ref{prop153_1}) We can rewrite the PDE \eqref{eq153_1} in a form that will remind us of the heat equation with non constant diffusion coefficient.
By changing variables $z=e^{y}$, we get $v(t,z)=\tilde{v}(t,y)$  
Write $\tilde{\mathbb{Q}}(t,y)=\mathbb{Q}(t,e^y)$ and define $\tilde{r}, \tilde{\theta},\tilde{\sigma}$ similarly. We get
$$\frac{\partial v}{\partial z}(t,z)=\frac{1}{z}\frac{\partial \tilde{v}}{\partial y}(t,y), \;\;  \;\; \frac{\partial^2 v}{\partial z^2}(t,z)=-\frac{1}{S^2}\frac{\partial \tilde{v}}{\partial y}(t,y)+\frac{1}{z^2}\frac{\partial^2 \tilde{v}}{\partial y^2}(t,y).$$
Then $\tilde{v}$ satisfies the PDE:
\begin{eqnarray}\label{eq153_3}
&& \frac{\partial \tilde{v}}{\partial t}(t,y)+\frac{\tilde{\sigma}^2(t,y)}{2}\frac{\partial^2 \tilde{v}}{\partial z^2}(t,y)+(\tilde{r}+p\tilde{\sigma}\tilde{\theta}-\frac{\tilde{\sigma}^2}{2})(t,y)\frac{\partial \tilde{v}}{\partial y} (t,y) \nonumber\\
&&+p(\g \tilde{r}+\frac{\g p\tilde{\theta}^2}{2}-\tilde{\mathbb{Q}})(t,y)\tilde{v}(t,y)+1=0 \\
&& \tilde{v}(T,z)=1. \nonumber
\end{eqnarray}

For the proof of Proposition \ref{prop153_1}, it suffices to invoke Theorem 4.6,  Chapter 6 of \cite{Fr1} to the above non degenerate linear parabolic PDE.  From $\tilde{v}$ we get $v$. 
The boundedness of $v$ away from zero yields when the following Feynman Kac's representation is used
$$
v(t,z)= \sE \bigg[\int_t^T  e^{\int_t^s  p(\g r+\frac{\g p \theta^2}{2}-\mathbb{Q})(u,\bar{S}(u)) du} ds + e^{\int_t^T  p(\g r+\frac{\g p \theta^2}{2}-\mathbb{Q})(u,\bar{S}(u)) du}| {\bar{S}(t)}=z  \bigg],
$$
where $\bar{S}(u)$ satisfies the SDE
\be
\bar{S}(u) = z +\int_t^u (r( v,\bar{S}(v))+p\sigma\theta(v,\bar{S}(v)))\bar{S}(v) dv +\int_t^u \sigma(v,\bar{S}(v))\bar{S}(v) dW(v).
\ee
To establish the boundedness of $z\frac{\partial v}{\partial z},$ notice that this is equivalent to the boundedness of $\frac{\partial \tilde{v}}{\partial y}.$ This
can be established by  a Feynman Kac's representation of the quotient
$$ \frac{\tilde{v}(t,y+h)- \tilde{v}(t,y) }{h }. $$
 Then the use of the mean value theorem combined with the finiteness of random variables moments entering the Feynman Kac's functional form yields $$ \left| \frac{\tilde{v}(t,y+h)- \tilde{v}(t,y) }{h } \right|\leq K, $$
for a constant $K$ independent of $h$ and $t.$ Then let $h\rightarrow 0$ to get the claim.
\ep

\proof (Theorem \ref{thm153_1}) Recall that $(\bar{\pi},\bar{c})$ is given by
 \be \bar{c}(t)=\frac{1}{v(t,S(t))} \; , \; \bar{\pi}(t)=p\frac{\theta(t, S(t))}{\sigma (t, S(t))}+\frac{S(t) \frac{\partial v}{\partial S} (t, S(t)) }{v(t,S(t))} \ee and the corresponding wealth process
$\bar{X}(s)=X^{\bar{\pi},\bar{c}}(s).$
By It$\hat{\mbox{o}}$'s lemma and using that $v$ solves \eqref{eq153_1} we get
\begin{eqnarray}
&&d\log (\bar{c}(t)\bar{X}(t)) =p( r+\frac{\theta^2}{2}-\mathbb{Q}) (t, S(t))dt+p\theta (t, S(t)) dW(t).
\end{eqnarray}
Therefore,
\be
\bar{c}(s)\bar{X}(s)=\bar{c}(t)\bar{X}(t) \exp\bigg(\int_t^s p( r+\frac{\theta^2}{2}-\mathbb{Q}) (u, S(u))du+\int_t^s p\theta (u, S(u)) dW(u)\bigg).
\ee

Note that  $\bar{c}(T)=\frac{1}{v(T,z)}=1$ yields

\begin{eqnarray*}
&&\mathbb{Q}(t,S(t))=F[\mathbb{Q}](t,S(t))=\frac{\sE_t \! \bigg[\int_t^T \!\! \frac{\partial h(t,s)}{\partial t} U_{\g}(\bar{c}(s)\bar{X}(s))ds+\frac{\partial h(t,T)}{\partial t} U_{\g}(\bar{c}(T)\bar{X}(T))\bigg]}{\sE_t \bigg[\int_t^T  h(t,s) U_{\g}(\bar{c}(s)\bar{X}(s))ds+h(t,T) U_{\g}(\bar{c}(T)\bar{X}(T))\bigg]}\\
&&\mathbb{Q}(t,S(t))=Q^{\bar{\pi},\bar{c}}(t).
\end{eqnarray*}

Next define
\be {V}(t,z,x)=\sE_t \bigg[\int_t^T  h(t,s) U_{\g}(\bar{c}(s)\bar{X}(s))ds+h(t,T) U_{\g}(\bar{c}(T)\bar{X}(T))\bigg].
\ee

It turns out that

$${V}(t,z,x)=Z(t,z)U_{\g}\big(\frac{x}{v(t,z)}\big), $$

where

$$Z(t,z)= \sE_t \bigg[\int_t^T  h(t,s) e^{\int_t^s  p\g( r+\frac{\theta^2}{2}-\mathbb{Q}) (u, S(u))du+p\g\theta (u, S(u)) dW(u)} ds+$$$$
+h(t,T)e^{\int_t^T  p\g( r+\frac{\theta^2}{2}-\mathbb{Q}) (u, S(u))du+p\g\theta (u, S(u)) dW(u)}\bigg]. $$
This is the Feyman Kac representation of  PDE \eqref{eq153_1}'s solution, and in light of uniqueness
$$ Z=v,$$
whence

 \be
{V}(t,z,x)= Z(t,z)U_{\g}\bigg(\frac{x}{v(t,z)}\bigg)=v(t,z)^{1-\g}U_{\g}(x).
\ee
A direct calculation gives:
$$\frac{\partial V}{\partial t}+\mathcal{A}^{\bar{\pi}, \bar{c}}{V} +U_{\g}(x \bar{c}) =Q^{\bar{\pi},\bar{c}}(t)v(t,z,x),$$
$$(\bar{\pi}, \bar{c})= \arg\max_{\pi,c}\{ \mathcal{A}^{\pi,c}{V} +U_{\g}(x c)\},$$
so ${V}$ satisfies 
\begin{equation}
\frac{\partial V}{\partial t}+\sup_{\pi,c}\{ \mathcal{A}^{\pi,c}{V} +U_{\g}(x c)\} =Q^{\bar{\pi},\bar{c}}(t){V}(t,z,x).
\end{equation}
Thus, $(V, \bar{\pi}, \bar{c})$ solves the extended HJB. Let us now turn to admissibility of $(\bar{\pi}, \bar{c}).$ The positivity of $\{ \bar{c}(t) \}_{\{0\leq t\leq T\}}$ and $\bar{X}(T)$ is straightforward. The strategy $(\bar{\pi}, \bar{c})$ is bounded by Proposition \ref{prop153_1}, and as such is admissible.
\ep

%
%
%

\proof (Proposition \ref{IND}) The Feyman-Kac representation of $\hat{v},$ yields

\begin{eqnarray}\label{eq153_2n}
\hat{v}(t,z)&=& \sE_t \bigg[\int_t^T  e^{\int_t^s  p(\g r+\frac{\g p \theta^2}{2}-\rho(0,u))(u,\bar{S}(u)) du} ds +  e^{\int_t^T  p(\g r+\frac{\g p \theta^2}{2}-\rho(0,u))(u,\bar{S}(u)) du}| {\bar{S}(t)}=z  \bigg], \nonumber \\
\end{eqnarray}
whence
$$ \hat{v}(t,z)=  \int_t^T A(t,s,z)(1+\delta_{T}(s))  e^{-\int_t^s \rho(0,u)du}  ds,$$
with
$$   A(t,s,z) =\sE_t \bigg[ e^{\int_t^s  p(\g r+\frac{\g p \theta^2}{2})(u,\bar{S}(u)) du} ds | {\bar{S}(t)}=z  \bigg],$$
and $\delta_{T}(s)$ the Dirac function.
Consequently 
$$ \frac{\partial\hat{v}}{\partial S} (t,z)=  \int_t^T A_{S}(t,s,z)(1+\delta_{T}(s))  e^{-\int_t^s \rho(0,u)du} ds. $$
 By the generalized mean value theorem 
 $$  \frac{\frac{\partial\hat{v}}{\partial z} (t,z)}{\hat{v}(t,z)}=\frac{A_{S}(t,s_0,S)  }{ A(t,s_0,S)  },$$
 for some $s_0 \in (t, T),$ hence the claim yields.

\ep

\subsection{ Numerical Scheme}

Consider the probability $\bar{\sP}$ with density
$$\frac{d\bar{\sP}}{d \sP}=\exp\bigg(\int_0^T p\g \theta(u, S(u)) dW(u) -\frac{1}{2}\int_0^T (p\g \theta(u, S(u)))^2 du\bigg),$$
and
\be
\mu_1 = r +\frac{p \theta^2}{2}.
\ee
The stock has the dynamics
\be
d S(u) = S(u) (r+p\sigma \theta)(u,S(u)) du +\sigma(u,S(u))d\bar{W}(u).
\ee
under $\bar{\sP}.$ Then

\be
\alpha(t,s,z)=\sE ^{\bar{\sP}}  \left[e^{\int_t^s p\g (\mu_1-\mathbb{Q})(u,S(u)) du} \bigg|  S(t)=z\right].
\ee

We discretize time and space at the points 

$$\{t_n = T-n \; dt \; , \; n=0, \cdots, N \}, \,\, \{ S_i = i \; dS \; , \; i=0, \cdots , M \}.$$

Let $Q_{n,i}$ be an approximation for $\mathbb{Q}(t_n, S_i).$ We start with $t_0=T,$ and set $Q_{0,i}=\frac{\frac{\partial h}{\partial t}(T,T)}{h(T,T)}.$ 

Next, use the relations
\be
\mathbb{Q}(t,z)=\frac{\int_t^T \frac{\partial h}{\partial t}(t,s)\alpha(t,s,z)ds+\frac{\partial h}{\partial t}(t,T)\alpha(t,T,z)}{\int_t^T h(t,s)\alpha(t,s,z)ds+h(t,T)\alpha(t,T,z)}, \ee
and approximate the integrals using a Riemann approximation and the conditional expectations through Monte Carlo simulations. This leads to

\be
Q_{n,i}=\frac{dt \sum_{j=0}^n  \frac{\partial h}{\partial t}(t_n,t_j)\alpha_{n,j,i}+\frac{\partial h}{\partial t}(t_n,t_0)\alpha_{n,0,i}}{dt \sum_{j=0}^n  h(t_n,t_j)\alpha_{n,j,i}+h(t_n,t_0)\alpha_{n,0,i}}. \ee
The above fixed point equation is calculated by starting with an initial guess for $Q_{n,i},$ then iterate the right hand side until the error is small enough.

\end{document}